\RequirePackage{fix-cm}
\documentclass{svjour3}                     

\smartqed  
\usepackage{graphicx}
\usepackage[a4paper,margin=1in]{geometry}
\usepackage[authoryear]{natbib}
\usepackage[dvipsnames]{xcolor} 
\usepackage{xspace}
\usepackage{listings}
\usepackage{subcaption}
\usepackage{comment}
\usepackage{tcolorbox}
\usepackage{amssymb}
\usepackage{amsmath}

\usepackage{tabularx}
\usepackage{colortbl}

\usepackage{url}
\usepackage{hyperref}

\usepackage{graphicx}

\usepackage{array}			
\usepackage{multirow}
\usepackage{tabularx}
\usepackage{booktabs} 
\usepackage{tikz}

\newcommand{\eg}{e.g.,~}							
\newcommand{\ie}{i.e.,~}							
\newcommand{\etal}{~et al.}					
\newcommand{\Fig}[1]{Figure~\ref{#1}}  			
\newcommand{\Table}[1]{Table~\ref{#1}}	    
\newcommand{\Sect}[1]{Section~\ref{#1}}	  

\newcommand{\up}{\textcolor{ForestGreen}{\(^\uparrow\)}}
\newcommand{\down}{\textcolor{red!100!black}{\(^\downarrow\)}}

\usepackage{xcolor}

\journalname{Empirical Software Engineering}
\begin{document}

\title{Balancing Usefulness and Naturalness: An LLM-based Curation Pipeline for Code Review Comments}

\author{\hspace{-0.18em}Oussama~Ben~Sghaier\hspace{-0.18em} \and
        \hspace{-0.18em}Martin~Weyssow\hspace{-0.18em} \and
        \hspace{-0.18em}Houari~Sahraoui}


\institute{
O. Ben Sghaier \at
School of Computing, Queen’s University, Kingston, ON, Canada \\
\email{oussama.sghaier@queensu.ca}
\and
M. Weyssow \at
Singapore Management University, Singapore \\
\email{mweyssow@smu.edu.sg}
\and
H. Sahraoui \at
DIRO, Université de Montréal, Montréal, QC, Canada \\
\email{sahraouh@iro.umontreal.ca}
}

\date{Accepted: 6 July 2026}

\maketitle

\begin{abstract}
Code review is a cornerstone of software development, where reviewers provide feedback through written comments to ensure code quality, maintainability, and correctness. The effectiveness of this process hinges on the quality of review comments: they must be clear, concise, actionable, and realistic enough to guide accurate code changes. As large language models (LLMs) gain traction in automating code review tasks, the utility of these systems is directly limited by the quality of the datasets on which they are trained. Unfortunately, existing code review datasets are often noisy, inconsistent, or poorly structured, which hinders the ability of LLMs to learn to generate accurate, helpful, and human-like review comments.

To overcome these limitations, we propose two different curation pipelines designed to improve both the quality and the utility of large-scale code review datasets. In the first pipeline, all review comments are systematically reformulated by an LLM to improve their clarity, conciseness, and civility while preserving their semantic intent. The curated dataset resulting from this approach, called \textbf{CuREV}, offers cleaner, higher-quality, and easier-to-learn-from comments that lead to measurable improvements in downstream automation tasks, namely review comment generation and code refinement. Building on this, we propose an improved pipeline, guided by high-quality exemplars, that enhances the realism and diversity of curated review comments. This method first separates the dataset into high-quality (``good'') and low-quality (``poor'') reviews, based on a systematic quality assessment using an evaluation framework. High-quality comments are preserved in their original form and further used as in-context exemplars to inspire the reformulation of low-quality comments. By varying the exemplars provided, the reformulated comments are not only clearer and more actionable but also exhibit a broader range of writing styles, making them more realistic and human-like. The resulting dataset, called \textbf{CuREV+}, thus combines improved quality and utility with enhanced diversity of review comments.

We evaluate both curated datasets using a comprehensive evaluation framework that assesses review comments along multiple quality dimensions (\eg clarity, conciseness, civility, nature) as well as their impact on downstream code review tasks compared to the original dataset. In addition, we analyze the diversity of the curated datasets, \emph{CuREV} and \emph{CuREV+}. Our results show that while both approaches significantly enhance review comment quality and improve the performance of automated code review tasks, \emph{CuREV+} provides a more diverse dataset, enriched with broader vocabulary, varied lexical choices, and distinct writing styles. These findings demonstrate that curating datasets for code review requires not only refining quality but also balancing standardization with diversity.  

\end{abstract}

\section{Introduction}
\label{sec:intro}

Code review is one of the most widely adopted practices in software development. It enables teams to detect defects early, improve maintainability, and transfer knowledge between developers~\citep{mcintosh2014impact, mcintosh2016empirical, ackerman1989software, morales2015code}. The process typically consists of peers inspecting the code and exchanging feedback in the form of written comments~\citep{fagan2002design, bavota2015four}. These comments are the cornerstone of the process: they document potential issues, suggest concrete improvements, and clarify the intent of the requested changes. The quality of these comments, whether they are clear, concise, relevant, and civil, directly influences the efficiency and outcomes of the code review process. Without high-quality review comments, developers risk overlooking bugs, misinterpreting design choices, or engaging in unproductive discussions~\citep{bacchelli2013expectations, sadowski2018modern}.  

As the volume and complexity of software projects grow, automating parts of the review process has become an appealing goal. Large language models (LLMs) and natural language processing (NLP) techniques have shown promising results in automatically generating review comments that identify code issues and suggest improvements~\citep{tufan2021towards, tufano2022using, li2022automating, li2022auger, sghaier2024improving}. Automated comment generation not only has the potential to reduce the cognitive load on reviewers but also to increase consistency and coverage in identifying problems. However, despite these advances, current systems still struggle to capture the nuance, clarity, and contextual awareness of human reviewers, often failing to generate review comments that are both accurate and useful~\citep{lu2025deepcrceval, li2022auger}.

A key limitation lies in the datasets used to train such models. Publicly available code review datasets, while large and diverse, are often mined in raw form from repositories and lack any curation~\citep{li2022automating, tufano2021towards, sghaier2023multi}. As a result, they contain a considerable amount of noise, including irrelevant or uncivil comments, vague or unclear phrasing, grammatical issues, and unstructured feedback~\citep{sghaier2025harnessing}. Training LLMs directly on such data not only reduces their learning efficiency but also risks amplifying undesirable behaviors, such as generating irrelevant, unclear, incivil, or overly verbose comments. Since the quality of language models is tightly coupled with the quality of their training data, curating code review datasets becomes a critical prerequisite for effective automation.  

In our previous work~\citep{sghaier2025harnessing}, we introduced \emph{CuREV}, a curated dataset created through a systematic reformulation pipeline designed to enhance the quality of code review comments. In this approach, each comment was reformulated by a large language model (LLM) to preserve its semantic intent while improving its clarity, conciseness, and civility. The resulting dataset offered standardized, high-quality review comments that were more consistent and easier for LLMs to learn from, leading to significant performance gains in downstream automation tasks such as review comment generation and code refinement. 

While this fully reformulated dataset raised overall quality, it also homogenized the dataset and reduced the natural variability of writing styles. By reformulating all comments, the dataset became more uniform, leading to the loss of natural stylistic variability of real-world review comments authored by developers with diverse backgrounds and communication styles. This led to comments that, although clean and high-quality, were more synthetic and less representative of natural developer interactions.  

To address this limitation, we propose an improved and more selective curation pipeline. This approach first separates the original dataset into high-quality (``good'') and low-quality (``poor'') comments, using our evaluation framework as the basis for classification. High-quality comments are preserved in their original form, thereby maintaining a natural, human-like core in the dataset. These good comments are then employed as in-context exemplars to inspire the reformulation of poor comments. By varying the exemplars provided, the reformulated comments gain stylistic diversity and lexical richness, resulting in a new curated dataset that we call \emph{CuREV+}. This dataset not only improves clarity and usefulness but also preserves diversity and realism, producing review comments that better reflect human-written styles.

In this paper, we evaluate \emph{CuREV+} against both the original dataset and \emph{CuREV}, using our evaluation framework that assesses review comments along several quality dimensions, including nature, relevance, clarity, conciseness, and civility. Beyond intrinsic quality, we also measure the impact of \emph{CuREV+} on downstream tasks, namely comment generation and code refinement, and compare it against both the original dataset and \emph{CuREV}. Finally, we analyze the diversity of review comments in \emph{CuREV+}, with respect to lexical richness, vocabulary distribution, and stylistic variation, contrasting it with the diversity offered by \emph{CuREV}.  

In summary, this paper makes the following contributions:  
\begin{itemize}
    \item We propose an improved approach for the systematic reformulation of review comments by introducing a new selective, in-context, exemplar-guided curation pipeline that produces the curated dataset \emph{CuREV+}.  
    \item We evaluate \emph{CuREV+} using our evaluation framework and compare it against both the original dataset and \emph{CuREV}, highlighting quality improvements across multiple dimensions.  
    \item We conduct an empirical study on the impact of \emph{CuREV+} on downstream code review tasks—namely, comment generation and code refinement—compared to the original dataset and \emph{CuREV}.  
    \item We provide a comparative analysis of the diversity of review comments in \emph{CuREV+} and \emph{CuREV}, demonstrating that \emph{CuREV+} produces a more diverse and human-like dataset.  
\end{itemize}

The remainder of this paper is organized as follows.
\Sect{sec:methodology} details our overall methodology for curating and evaluating code review datasets and answering the different research questions. 
\Sect{sec:initdata} presents our proposed evaluation framework and the initial quality assessment of the original code review dataset.
\Sect{sec:curdata} introduces the two curation pipelines that yield \emph{CuREV} and \emph{CuREV+}, whose quality is then evaluated using our evaluation framework.
\Sect{sec:diversity} examines the diversity of both curated datasets.
Finally, \Sect{sec:analysis} conducts a comparative study on the impact of the original and curated datasets on downstream tasks, namely code review comment generation and code refinement.
\Sect{sec:threats} discusses threats to validity, \Sect{sec:related} reviews related work, and \Sect{sec:conclusion} concludes the paper.

\paragraph{Clarification on Naturalness}
In this context, we use the term \emph{naturalness} to refer strictly to the stylistic authenticity and human-likeness of review comments---i.e., the extent to which curated comments preserve the natural variation and informal conventions of human-written reviews. This is distinct from the statistical naturalness of source code studied by \citet{hindle2012naturalness}, which primarily concerns the repetitive and predictable nature of code tokens.

\section{Methodology}
\label{sec:methodology}

\begin{figure*}[!h]
    \centering
    \includegraphics[width=1\textwidth]{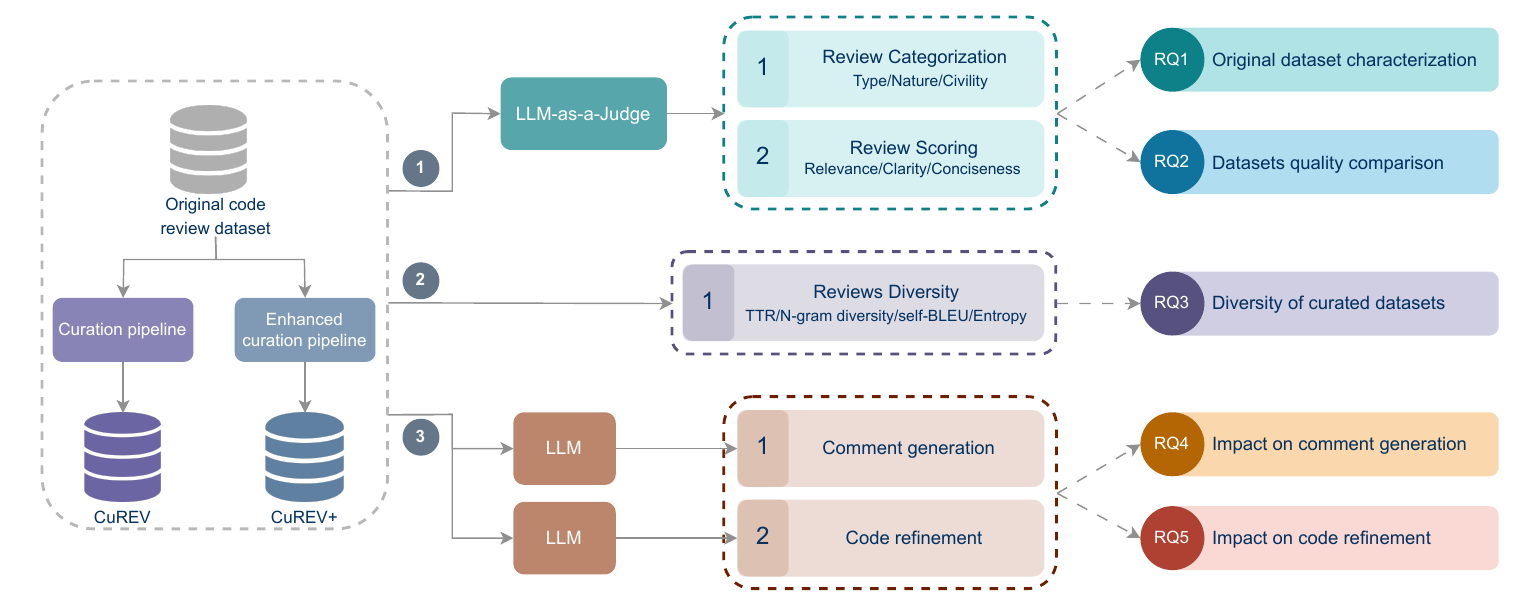}
    \caption{Overview of the proposed methodology. We begin by assessing the quality of the original code review dataset (\emph{RQ1}) using an evaluation framework and an LLM-as-a-Judge setup. Two curation pipelines are then applied to produce \emph{CuREV} and \emph{CuREV+}. We evaluate their quality (\emph{RQ2}), diversity (\emph{RQ3}), and their impact on two downstream tasks: comment generation (\emph{RQ4}) and code refinement (\emph{RQ5}).}
    \label{fig:methodology}
    \vspace{-.5em}
\end{figure*}

This section details the methodology followed to assess and enhance the quality of an existing code review dataset and to evaluate the impact on downstream software engineering tasks. Our process, illustrated in \Fig{fig:methodology}, is structured around five research questions (RQs) that guide our empirical investigation.

\begin{itemize}
    \item \textbf{RQ1:} \emph{What are the main characteristics and quality concerns in the original code review dataset?}
    \item \textbf{RQ2:} \emph{How can our curation pipelines improve dataset quality?}
    \item \textbf{RQ3:} \emph{How do the curated datasets, \emph{CuREV} and \emph{CuREV+}, compare in terms of linguistic and stylistic diversity?}
    \item \textbf{RQ4:} \emph{What is the impact of the curated datasets (\emph{CuREV} and \emph{CuREV+}) compared to the original dataset on the performance of LLMs for automated comment generation?}
    \item \textbf{RQ5:} \emph{What is the impact of the curated datasets (\emph{CuREV} and \emph{CuREV+}) compared to the original dataset on the usefulness and effectiveness of LLMs for automated code refinement?}

\end{itemize}

\subsection{Original Dataset Characterization (RQ1)}

\begin{figure}[!t]
    \centering
    \includegraphics[width=\linewidth]{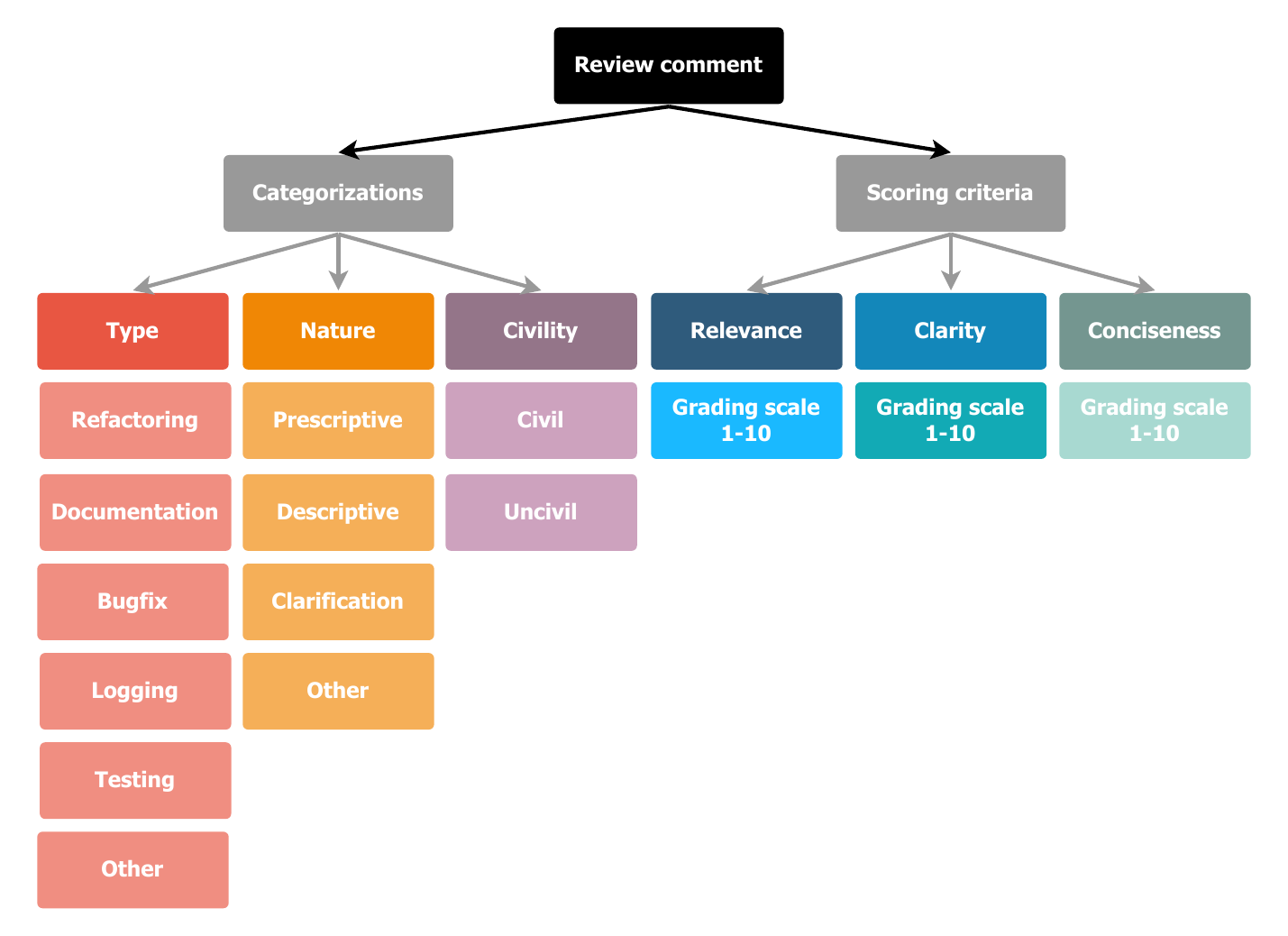}
    \caption{Evaluation framework for categorizing and scoring review comments.}
    \label{fig:eval_framework}
\end{figure}

We begin with a large-scale dataset of code reviews containing old code and code change pairs, along with the associated review comments \citep{li2022automating}. To assess its initial quality, we employ an evaluation framework (\Fig{fig:eval_framework}) that classifies review comments along three categorical dimensions (\ie \emph{type}, \emph{nature}, and \emph{civility}) and scores them across three quality criteria (\ie \emph{relevance}, \emph{clarity}, and \emph{conciseness}).  

Using this framework, we apply \emph{Llama-3.1-70B} in an \emph{LLM-as-a-Judge} setup to automatically evaluate each review comment, generating both its categorical labels and corresponding scores. This step allows us to identify weaknesses and inconsistencies in the original dataset and to serve as a baseline for subsequent curation and comparison.

\subsection{Dataset Curation and Quality Comparison (RQ2)}

Building upon the insights from Step~1, we introduce two complementary curation pipelines designed to improve the overall quality of the original code review dataset:

\begin{itemize}
    \item \textbf{CuREV:} a filtering and reformulation pipeline that removes irrelevant and uses the LLM to rewrite the retained comments for better clarity, conciseness, and civility while preserving content and intent.
    \item \textbf{CuREV+:} an enhanced variant that combines selective filtering and targeted reformulation of low-quality comments, guided by the same quality attributes (\ie clarity, conciseness, and civility) and inspired by high-quality examples used as in-context exemplars, while preserving the original content and intent.
\end{itemize}

The outputs of both pipelines are evaluated using the same evaluation framework employed in Step~1 (\Fig{fig:eval_framework}), enabling a consistent comparison with the original dataset in terms of categorical distributions and quality scores.

\subsection{Diversity Analysis (RQ3)}

To assess linguistic richness and stylistic variety, we measure and compare the diversity of review comments across the two curated datasets (CuREV and CuREV+). We use several complementary metrics: \emph{type–token ratio (TTR)}, \emph{$n$-gram diversity} (bi- and trigram), \emph{self-BLEU}, and \emph{entropy}. These indicators capture redundancy, lexical variability, and overall expressiveness, helping us determine whether the curation processes led to oversimplification and excessive uniformity or improved balance between quality and diversity.

\subsection{Impact on Comment Generation (RQ4)}

To examine how dataset curation affects LLM training, we fine-tune three separate models on each dataset (original dataset, CuREV, and CuREV+) for the task of \emph{review comment generation}. Given a code change, the model is trained to predict the review comment. We evaluate the generated comments using the BLEU score to compare how well each dataset supports the learning of precise review comments.

\subsection{Impact on Code Refinement (RQ5)}

Finally, we evaluate the practical usefulness of the review comments in guiding code improvements. Three versions of the same base model are provided with either the original or curated comments to perform \emph{code refinement} (\ie implement the suggested review changes). We measure the correctness of the resulting code using CodeBLEU and Exact Match (EM). This step quantifies the end-to-end benefit of dataset curation, linking improved comment quality to more effective code refinement outcomes.

\subsection{Terminology and Operational Definitions}
To ensure clarity throughout our evaluation and curation framework, we establish the following operational definitions:
\begin{itemize}
    \item \textbf{Quality:} The degree to which a review comment is clear, concise, relevant, civil, and actionable.
    \item \textbf{Naturalness:} The stylistic authenticity and human-likeness of review comments (distinct from code naturalness per \citet{hindle2012naturalness}).
    \item \textbf{Usefulness:} The ability of a review comment to successfully guide developers toward correct code changes.
    \item \textbf{Diversity:} The lexical and stylistic variety across review comments collectively within a dataset.
    \item \textbf{Curation:} The systematic automated process of improving dataset quality through iterative filtering, classification, and reformulation of review comments.
\end{itemize}

\section{Quality Assessment and Characterization of Existing Code Review Dataset}
\label{sec:initdata}

In this section, we present the evaluation framework, detail its application in assessing the existing code review dataset, and then present the results.

\subsection{Dataset Selection}
We employ the code review dataset introduced in~\citep{li2022automating}, which is the largest publicly available dataset for code reviews. This dataset has been widely adopted in several works~\citep{li2022automating, ben2024improving, sghaier2023unity, lu2023llama} for automating code review tasks.
The dataset is multilingual, covering nine programming languages, and consists of $176,613$ samples. \Table{tab:dataset_stats} shows the distribution of the dataset across the different programming languages.
The dataset includes features such as the code changes submitted for review and the review comment provided by the reviewer assigned to the pull request. 


\begin{table}[!t]
  \centering
  \caption{Dataset distribution over programming languages.}
  \label{tab:dataset_stats}
  \begin{tabular}{{>{\centering\arraybackslash}p{3cm}}*{1}{r}}
    \toprule
    \textbf{Programming Language} & \textbf{\# Samples} \\
    \midrule
    PHP & $9,984$ \\
    Ruby & $6,713$ \\
    C\# & $17,085$ \\
    C & $4,108$ \\
    Java & $35,671$ \\
    Python & $36,382$ \\
    C++ & $15,944$ \\
    Go & $36,123$ \\
    JS & $14,603$ \\
    \midrule
    \textbf{Total} & $176,613$\\
    \bottomrule
  \end{tabular}
  \vspace{-1em}
\end{table}

\subsection{Evaluation Framework}

In this section, we introduce the evaluation framework used to assess the quality and characteristics of the code reviews in existing datasets. 
As shown in \Fig{fig:eval_framework}, this framework consists of three key categorizations, along with a scoring system to evaluate the clarity, relevance, and conciseness of each review comment. These categories and criteria aim to provide a comprehensive analysis of the reviews and enable us to identify areas for improvement in the dataset.

\Table{tab:categories} presents the categorization framework for code reviews, which is used to classify each review comment based on three key aspects: \emph{Type}, \emph{Nature}, and \emph{Civility}. The \emph{Type} category identifies the primary focus of the review, such as refactoring, bugfixes, or documentation~\citep{tufano2024code}. The \emph{Nature} category assesses the intent behind the comment, categorizing it as prescriptive, descriptive, or clarifying. Finally, the \emph{Civility} category evaluates the tone of the comment; whether it is civil or uncivil~\citep{rahman2024words}. 
Note that \emph{Type} and \emph{Nature} categories are multi-labeled. That is, a single review comment may address multiple types of issues (\eg documentation and testing) or have a mixed nature (\eg descriptive and requests clarifications).
This categorization provides a structured approach for systematically evaluating the dataset, offering a detailed breakdown of the content and intent of review comments.

\begin{table}[!t]
  \centering
  \caption{Categorization framework for reviews comments.}
  \label{tab:categories}
  \begin{tabularx}{0.7\linewidth}{llX}
    \toprule
    \textbf{Category} & \textbf{Subcategory} & \textbf{Description} \\
    \midrule
    \textbf{Type} & Refactoring & Suggestions to improve code structure \\
                  & Bugfix & Identifies and suggests fixes for bugs \\
                  & Testing & Comments related to test cases \\
                  & Logging & Suggestions for logging practices \\
                  & Documentation & Recommendations for documentation changes \\
                  & Other & Any other type of comment \\
    \midrule
    \textbf{Nature} & Prescriptive & Provides specific actions or recommendations \\
                    & Descriptive & Describes a situation without suggesting changes \\
                    & Clarification & Requests or provides clarification \\
                    & Other & Any other nature of comment \\
    \midrule
    \textbf{Civility} & Civil & Respectful and professional tone \\
                      & Uncivil & Disrespectful or inappropriate tone \\
    \bottomrule
  \end{tabularx}
\end{table}

\Table{tab:criteria} outlines the scoring criteria used to assess the quality of reviews in three dimensions: \emph{clarity}, \emph{relevance}, and \emph{conciseness}~\citep{rani2023decade, haouari2011good}. Each criterion is scored on a scale from 1 to 10. \emph{Clarity} measures how effectively the comment communicates its message, with higher scores reflecting clearer communication. \emph{Relevance} evaluates how pertinent the comment is to the code change, and \emph{Conciseness} assesses whether the comment is brief and to the point without unnecessary details. These criteria provide a structured way to evaluate the quality and effectiveness of each comment, ensuring a comprehensive assessment of the dataset.

\begin{table}[!t]
  \centering
  \caption{Scoring criteria for review comments.}
  \label{tab:criteria}
  \begin{tabularx}{\linewidth}{lX}
    \toprule
    \textbf{Criterion} & \textbf{Description} \\
    \midrule
    \textbf{Clarity (1--10)} & Assesses how clearly the review comment communicates its message. A score of 1 indicates very unclear, and 10 indicates very clear communication. \\
    \textbf{Relevance (1--10)} & Evaluates the extent to which the review comment is pertinent to the code change. A score of 1 means the comment is completely irrelevant, while a score of 10 means it is highly relevant. \\
    \textbf{Conciseness (1--10)} & Measures the brevity and efficiency of the review comment, ensuring it conveys the necessary information without unnecessary elaboration. A score of 1 indicates too verbose, and 10 indicates concise and to the point. \\
    \bottomrule
  \end{tabularx}
  \vspace{-1em}
\end{table}

This framework enables a structured analysis of the reviews, allowing us to assess their quality in a granular manner and providing the foundation for identifying trends, patterns, and areas for improvement within the dataset.

\subsection{LLM-as-a-Judge}

To implement the evaluation framework on the code review dataset, we leverage an LLM, specifically \emph{Llama-3.1-70B-Instruct}, to act as annotator for the various categories and criteria. Due to the substantial size of the dataset (176,613 samples), manual evaluation is infeasible. Drawing from findings in the literature~\citep{zheng2023judging, zhuo2023ice, chang2024survey}, we assume that a highly capable LLM can serve as a reliable substitute for human evaluators, accurately assessing review comments across the defined framework.
This claim is supported by previous research demonstrating that highly capable LLMs closely align with human judgments, achieving agreement rates comparable to human-to-human agreement~\citep{zheng2024judging, li2023alpacaeval}. Furthermore, LLMs have been recently exploited as judges for different software engineering tasks~\citep{zhuo2023ice, weyssow2024codeultrafeedback}, thereby justifying their use as annotators in this work.

For conciseness, we show an excerpt of the prompt \footnote{The full prompt is available in the \href{https://github.com/AI4CodeReview/CuREV}{replication package}.} used to evaluate review comments in \Table{tab:eval_promp}.
Initially, the LLM is instructed to generate what it considers an ideal review comment based on the provided code changes. 
Following this, the LLM is asked to evaluate the given review comment according to the different categories and criteria. 
The generated review serves as an implicit reference during this evaluation. 
This process is repeated across the entire code review dataset.

The prompt was chosen through multiple refinement iterations, following a trial-and-error approach based on observations. Initially, we designed a preliminary prompt and tested it on a curated set of manually selected examples. Based on the identified issues, we refined the prompt to enhance its effectiveness. This iterative process was informed by techniques from the literature that demonstrated effectiveness in similar contexts~\citep{weyssow2024codeultrafeedback}. 
For instance, one challenge encountered was the tendency of LLM to generate unstructured and verbose outputs in inconsistent formats, which hindered automated parsing and extraction of relevant information. To mitigate this issue, the prompt was augmented with explicit examples of the desired output format, thereby encouraging the LLM to produce more structured and consistent results.
Another observed limitation was the propensity of LLM to assign high scores (\eg 9 or 10) across all evaluation criteria, resulting in a lack of meaningful differentiation. To address this, detailed descriptions of each criterion were incorporated into the prompt, and the LLM was instructed to generate an exemplary review as a reference. This adjustment facilitated a more balanced and discriminative grading distribution. Additionally, we asked the LLM to provide a rationale for each evaluation, effectively introducing an implicit chain-of-thought process. This approach not only improved the explainability and justification of the assigned grades but also enhanced the overall reliability and quality of the evaluations.

\begin{table}[!t]
\centering
\caption{Excerpt of the prompt for comments evaluation. While this excerpt highlights conciseness for spatial reasons, the full prompt formally evaluates all six dimensions: Type, Nature, Civility, Relevance, Clarity, and Conciseness.}
\vspace{-.5em}
\label{tab:eval_promp}
\begin{tabularx}{\linewidth}{X}
\toprule

\textbf{\#\#\# Code review comment generation}

 Generate a review comment that you consider perfect for the code change without considering the given input comment. A review comment should highlight the main issues, improvements, or suggestions for the code changes. The generated review comment should be concise, relevant, clear, useful, and complete. \\
\vspace{.3em}

\textbf{\#\#\# Code review comment assessment}

Then, evaluate and categorize only the given review comment, written by a reviewer, based on the below criteria.
You can use the generated review comment as a reference to evaluate the given review comment.
Note that multiple labels are allowed for the categories "Type" and "Nature". \\
\\
\textsc{\textbf{1. Type:}} Categorize the review according to the type of issue it addresses: Refactoring, Bugfix, Testing, Logging, Documentation, Other.
\vspace{.3em}

\textsc{\textbf{2. Nature:}} Specify the nature of the review according to these categories:\\
- \textit{Descriptive:} describe what the reviewer observes without explicitly suggesting specific actions.\\
- \textit{Prescriptive:} suggest or request specific actions on the code.\\
- \textit{Clarification:} request explanation or further information to better understand the code changes.\\
- \textit{Other:} for comments that do not fit the previous categories.

\hspace{0.5\linewidth}\vdots

\textsc{\textbf{4. Conciseness:}} Assess how effectively the comment conveys its message using the fewest necessary words while remaining fully informative. A concise comment should be completely brief but informative, avoiding unnecessary details, repetition, or verbosity. Use a 1-to-10 rating scale.

\hspace{0.5\linewidth}\vdots

\textbf{\#\#\# Given review comment}\\
\{\textit{review\_comment}\}\\

\textbf{\#\#\# Code changes}\\
\{\textit{code\_diff}\}\\

\bottomrule
\end{tabularx}
\end{table}

\subsection{Sanity Check}
To ensure the reliability of the judgments made by \emph{Llama-3.1-70B-Instruct}, we conducted a sanity check involving human assessments. A random sample of $100$ review comments was manually evaluated by two authors according to the evaluation schema defined in \Fig{fig:eval_framework}. This dual-author assessment aimed to mitigate individual biases and provide a robust baseline for comparison. Conflicts between the annotators were carefully resolved through discussion to reach a consensus on each sample's evaluation.


This manual assessment provides a human baseline against which we can compare the LLM judging performance. To measure the agreement between LLM and human judgments, we used \emph{Cohen's kappa}, a statistical measure that accounts for the degree of agreement beyond chance~\citep{mchugh2012interrater}. Our analysis showed \emph{perfect agreement} for the \emph{civility} category ($1$), \emph{near-perfect agreement} for \emph{type} ($0.88$) and \emph{nature} ($0.82$) categories. For criteria, we observed \emph{near-perfect agreement} for \emph{relevance} ($0.85$), \emph{substantial agreement} for \emph{Conciseness} ($0.76$) and \emph{Clarity} ($0.64$). These values provide strong evidence of the LLM capacity to make reliable judgments.

Furthermore, since evaluating properties such as relevance, clarity, and conciseness logically falls on a 1--10 ordinal scale, we extended our evaluation by computing the quadratic weighted kappa, which formally accounts for the varying magnitude of disagreements.
\begin{table}[!htbp]
\centering
\caption{Inter-rater agreement between human annotators and the LLM judge using standard Cohen's kappa and quadratic weighted kappa.}
\label{tab:kappa}
\begin{tabular}{llcc}
\toprule
\textbf{Category/Criterion} & \textbf{Cohen's $\kappa$} & \textbf{Weighted $\kappa$ (quadratic)} & \textbf{Interpretation} \\
\midrule
Civility & $1.00$ & $1.00$ & Perfect \\
Type & $0.88$ & $0.88$ & Near-perfect \\
Nature & $0.82$ & $0.82$ & Near-perfect \\
Relevance & $0.85$ & $0.99$ & Perfect \\
Conciseness & $0.76$ & $0.97$ & Perfect \\
Clarity & $0.64$ & $0.97$ & Perfect \\
\bottomrule
\end{tabular}
\end{table}

As expected, the weighted kappa values for the ordinal criteria (Relevance, Conciseness, Clarity) are substantially higher than the unweighted values. This mathematically confirms that the disagreements on these criteria were predominantly minor integer steps, translating to near-perfect true agreement when correctly accounting for the ordinal nature of the scale.

This outcome supports existing literature suggesting that LLMs possess the capability to evaluate review comments with accuracy comparable to human reviewers~\citep{zheng2024judging, li2023alpacaeval}. Furthermore, it underscores the reliability of LLMs in the specific context of judging review comments.

\subsection{Multi-LLM Validation and Reliability}
To robustly mitigate any risk of self-evaluation bias and to confirm the structural integrity of our curation methodology, we conducted a comprehensive cross-validation scenario involving two additional, architecturally independent, state-of-the-art LLMs: MiniMax-M2.5 and GPT-5-nano. 

Both independent models evaluated a highly curated consensus subset ($N=1,668$ successfully paired evaluations) using the identical prompt logic. A cross-agreement correlation analysis mathematically validates the reliability of our framework.

\begin{table}[!t]

\centering
\caption{Cross-architecture agreement (Spearman Rank Correlation) across independent state-of-the-art LLMs evaluated on the highly curated consensus subset (N = 1,668).}
\label{tab:agreement_multi_llm}
\begin{tabular}{lccc}
\toprule
\textbf{Metric Dimension} & \textbf{Llama-3.1-70B} & \textbf{MiniMax-M2.5} & \textbf{GPT-5-nano} \\
\midrule
\textbf{Relevance} & Base Reference & $\rho = 0.582$ ($p < .001$) & $\rho = 0.601$ ($p < .001$) \\
\textbf{Clarity} & Base Reference & $\rho = 0.501$ ($p < .001$) & $\rho = 0.503$ ($p < .001$) \\
\textbf{Conciseness} & Base Reference & $\rho = 0.334$ ($p < .001$) & $\rho = 0.340$ ($p < .001$) \\
\midrule
\textbf{Average Quality} & \textbf{--} & \textbf{$\rho = 0.472$} & \textbf{$\rho = 0.481$} \\
\bottomrule
\end{tabular}
\end{table}

 Across entirely different proprietary and open-weight model architectures, the LLMs consistently demonstrated strong, statistically significant monotonic correlation regarding subjective evaluation scores: Relevance (Spearman $\rho \approx 0.601, p < 0.001$), Clarity ($\rho \approx 0.503, p < 0.001$), and Conciseness ($\rho \approx 0.340, p < 0.001$). 

Given the inherent subjectivity involved in grading varying linguistic nuances on a 1-to-10 scale, achieving $\sim 0.60$ agreement across independent LLM pipelines is a strong indicator of prompt stability. This robust multi-evaluator agreement addresses the concern of single-model self-evaluation bias and confirms that our automated curation framework assesses review qualities consistently and objectively.

\subsection{Categorization Results}

\begin{figure*}
    \centering
    \includegraphics[width=\linewidth]{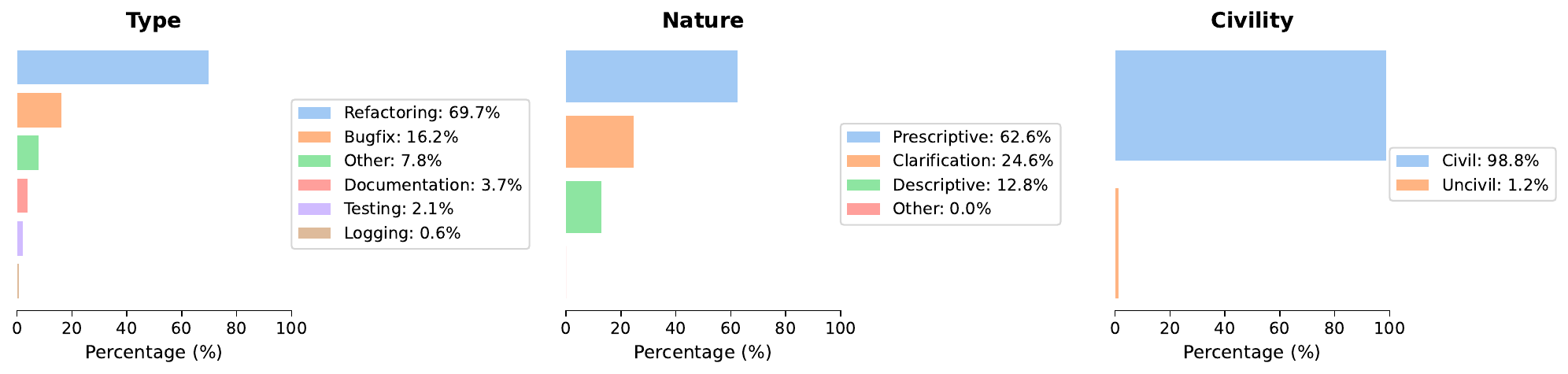}
    \caption{Distribution of the different categories across the original dataset.}
    \label{fig:initinal_categ_dist}
    \vspace{-1em}
\end{figure*}

\Fig{fig:initinal_categ_dist} presents the results of the different categorizations according to the \emph{type}, \emph{nature}, and \emph{civility} of the comments. The findings provide valuable insights into the original dataset characteristics and highlight potential areas of improvements.

\paragraph*{\textbf{Type of review comments\\}}

The majority of the review comments in the dataset fall under the \emph{Refactoring} category, comprising $80.07\%$ of the total. This indicates that much of the feedback is centered on restructuring the code to enhance its quality and performance. The next most significant category is \emph{Bugfix} at $18.60\%$, underscoring the importance of addressing functional issues within the code to avoid bugs.

Other subcategories, such as \emph{Documentation} ($4.21\%$), \emph{Testing} ($2.42\%$), and \emph{Logging} ($0.65\%$), are relatively less frequent. This suggests that these areas receive less attention in code review feedback, potentially due to implicit understanding among developers or prioritization of other aspects of the code. The \emph{Other} category, which makes up $8.97\%$ of the comments, encompasses miscellaneous feedback types that do not fall into the primary classifications, \eg security.

The focus on \emph{Refactoring} and \emph{Bugfix} comments align with the primary objectives of code review, \ie preserving code quality and preventing bugs. However, the lower representation of \emph{Documentation}, \emph{Testing}, and \emph{Logging} comments suggests potential gaps in the comprehensiveness of code review practices. Addressing this imbalance presents an opportunity for enhancing training data to support more well-rounded feedback generation. This finding can guide researchers in balancing their datasets according to the \emph{type} of review comment, ultimately leading to more varied and comprehensive model-generated feedback.

\paragraph*{\textbf{Nature of review comments\\}}

Review comments categorized as \emph{Prescriptive}, which provide direct guidance or specific instructions, dominate the dataset at $62.6\%$. This reflects a strong focus on actionable feedback, aiding developers in making precise code changes. \emph{Clarification} comments make up $24.6\%$, indicating a notable effort to ensure code understanding among team members. \emph{Descriptive} comments, which explain aspects of the code without providing explicit recommendations, account for $12.8\%$.
The \emph{Other} category is negligible at $0.01\%$, showing that most comments fit well into defined subcategories.

The prevalence of \emph{prescreptive} comments highlights a code review culture geared towards providing direct solutions and explicit feedback. The significant presence of \emph{Clarification} comments suggests that maintaining clear communication and understanding is a priority during the review process.

\paragraph*{\textbf{Civility of review comments\\}}

Most comments in the dataset are categorized as \emph{Civil}, making up $98.77\%$ of the total. This suggests that the code review process is generally conducted professionally and constructively. However, there is still a portion ($1.23\%$) of \emph{Uncivil} comments present in the dataset. These comments often contain harsh or inappropriate language that could negatively influence a model learning process.
Training on such comments risks introducing undesirable patterns, potentially leading to the model generating inappropriate language in downstream tasks.

To mitigate this, it is crucial to either curate these reviews to remove harsh language or exclude them altogether to prevent models from learning from these undesirable examples and reinforcing negative behavior.

\subsection{Scoring Criteria Results}

\begin{figure}
    \centering
    \includegraphics[width=.7\linewidth]{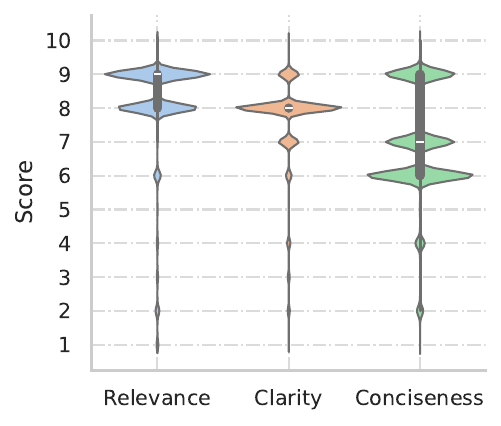}
    \caption{Distribution of scoring criteria on the original dataset.}
    \label{fig:init_scoring_dist}
\end{figure}

\Fig{fig:init_scoring_dist} depicts scoring criteria distribution across the original dataset. \Table{tab:init_categories_distribution} provides a summary of the average values for each scoring criterion within the different categories.

\begin{table}[!t]
  \centering
  \caption{Average values of the scoring criteria per category across the original dataset.}
  \label{tab:init_categories_distribution}
  \begin{tabular}{>{\centering\arraybackslash}p{1cm} >{\centering\arraybackslash}p{3cm}*{3}{c}}
    \toprule
    \textbf{Category} & \textbf{Subcategory} & \textbf{Relevance} & \textbf{Clarity} & \textbf{Conciseness} \\
    \midrule
    \multirow{6}{*}{\textbf{Type}} & Refactoring & $8.32$ & $7.79$ & $6.99$ \\
    & Bugfix & $8.53$ & $7.74$ & $6.84$ \\
    & Testing & $8.42$ & $7.92$ & $6.97$ \\
    & Logging & $8.43$ & $7.84$ & $6.85$ \\
    & Documentation & $8.33$ & $7.62$ & $6.72$ \\
    & Other & $7.02$ & $6.86$ & $5.90$ \\
    \midrule
    \multirow{4}{*}{\textbf{Nature}} & Descriptive & $7.14$ & $6.63$ & $5.61$ \\
    & Prescriptive & $8.52$ & $7.95$ & $7.19$ \\
    & Clarification & $8.29$ & $7.57$ & $6.66$ \\
    & Other & $4.24$ & $4.40$ & $4.12$ \\
    \midrule
    \multirow{2}{*}{\textbf{Civility}} & Civil & $8.26$ & $7.75$ & $6.93$ \\
    & Uncivil & $5.60$ & $4.34$ & $4.34$ \\
    \midrule
    \multirow{1}{*}{\textbf{Average}} & -- & $8.23$ & $6.89$ & $7.71$ \\
    \bottomrule
  \end{tabular}
\end{table}

\paragraph*{\textbf{Relevance of review comments\\}}
The highest relevance scores are found in the \emph{Bugfix} ($8.53$) and \emph{Prescriptive} ($8.52$) subcategories. The lowest relevance scores are observed in the \emph{Other} subcategories of both the \emph{Type} ($7.02$) and \emph{Nature} ($4.24$) categories, as well as in \emph{Uncivil} comments ($5.60$).
The low relevance of these subcategories indicates that such comments may often lack the focus or constructive value necessary for high-quality code reviews. 

The relevance distribution, illustrated in \Fig{fig:init_scoring_dist}, indicates that while the majority of comments are relevant, a portion of irrelevant comments persists, diminishing the overall quality of the dataset. These irrelevant comments can negatively influence language models' performance by introducing unhelpful patterns. 
This study can support the filtering of irrelevant comments to ensure a high-quality dataset, as maintaining data relevance is crucial for effective model training and downstream task performance.

\paragraph*{\textbf{Clarity of review comments\\}}

Clarity scores across the dataset are varied, with the highest scores found in the \emph{Prescriptive} $(7.95$), \emph{Testing} ($7.92$), and \emph{Refactoring} ($7.79$) subcategories, while the lowest clarity scores appear in the \emph{Other} subcategories ($4.40$) and \emph{Uncivil} comments ($4.34$).

The data indicates that \emph{Prescriptive} comments, which provide direct and specific guidance, are not only relevant but also clear, making them highly effective in guiding developers. High clarity scores in the \emph{Testing} and \emph{Refactoring} categories further suggest that comments in these areas are typically well-understood. In contrast, comments categorized as \emph{Other} and \emph{Uncivil} exhibit significantly lower clarity, which can lead to misunderstandings and inefficiencies during the code review process. The low clarity score (average = $6.89$) underscores the importance of enhancing the clarity of review comments to improve dataset quality, ultimately supporting the training of language models to produce clear and precise feedback that developers can easily comprehend and act upon.

\paragraph*{\textbf{Conciseness of review comments\\}}
The analysis of conciseness scores reveals that the highest average scores are found in the \emph{Prescriptive} ($7.19$), \emph{Refactoring} ($6.99$), and \emph{Testing} ($6.97$) subcategories. The lowest conciseness scores are observed in \emph{Other} subcategories for both \emph{Type} ($5.90$) and \emph{Nature} ($4.12$), as well as in \emph{Uncivil} comments ($4.34$).

The high conciseness scores in \emph{Prescriptive} comments indicate that these types of feedback tend to be not only clear and relevant but also succinct, contributing to more efficient communication. 
The lower conciseness scores suggest that these comments may be more verbose or include unnecessary language, reducing their effectiveness.

As shown in \Fig{fig:init_scoring_dist}, a significant portion of the dataset consists of comments that lack conciseness. This suggests an opportunity for enhancing the quality of the dataset by refining review comments to be more succinct, removing unnecessary elements, and conveying the message more efficiently. This improvement would enable language models to focus on the core content of the comments, emphasizing essential information over extraneous details.

\begin{center}
\begin{tcolorbox}[
    colback=teal!5!white,
    colframe=teal!60!black,
    left=3mm,
    overlay={\draw[teal!80!black, line width=3pt] (frame.south west)--(frame.north west);},
    title=\textbf{Answer to RQ1 (Quality of the Original Dataset)},
    fonttitle=\bfseries,
    boxrule=0pt,
    arc=2pt,
    before skip=5pt,
    after skip=5pt
]
The code reviews dataset is characterized by a strong focus on \emph{refactoring} ($80.07\%$) and \emph{bugfix} ($18.60\%$). Most of the comments are \emph{prescriptive} ($62.6\%$) providing direct and actionable suggestions to developers.
However, the presence of uncivil, lengthy, unclear, and irrelevant comments highlights areas for improvement to enhance dataset quality.
\end{tcolorbox}
\end{center}

\paragraph{Category Provenance and Civility Ontology}
The categories used in our evaluation framework are derived from established literature. The \emph{Type} categories (Refactoring, Bugfix, Testing, Logging, Documentation, Other) are derived from the foundational taxonomy of code review comments established by \citet{tufano2024code}. The \emph{Civility} dimension is based on the comprehensive framework proposed by \citet{rahman2024words}, who studied toxic and uncivil language in software engineering. In our context, an \emph{uncivil} comment is defined as containing harsh language, sarcasm, passive-aggressive tones, or personal attacks, while a \emph{civil} comment is respectful and constructive.

\paragraph{LLM Calibration and Sample Justification}
To properly calibrate the LLM judge on the 1-to-10 ordinal scale, the evaluation prompt included detailed categorical boundary descriptions. By instructing the model to generate an exemplary review as a reference baseline, we forced the model to calibrate its comparative assessment, successfully mitigating the tendency of LLMs to assign uniformly high scores. 
Furthermore, evaluating 100 samples from the dataset population provides a statistically rigorous 95\% confidence interval with an approximate $\pm$9.8\% margin of error, aligning with established practices for structural LLM-as-a-judge validations \citep{zheng2024judging}. Disagreements during human validation were resolved through structured multi-annotator review sessions focusing on the code context.

\section{Curated Code Review Datasets: CuREV and CuREV+}
\label{sec:curdata}

\begin{figure*}[!t]
    \centering
    \includegraphics[width=1\linewidth]{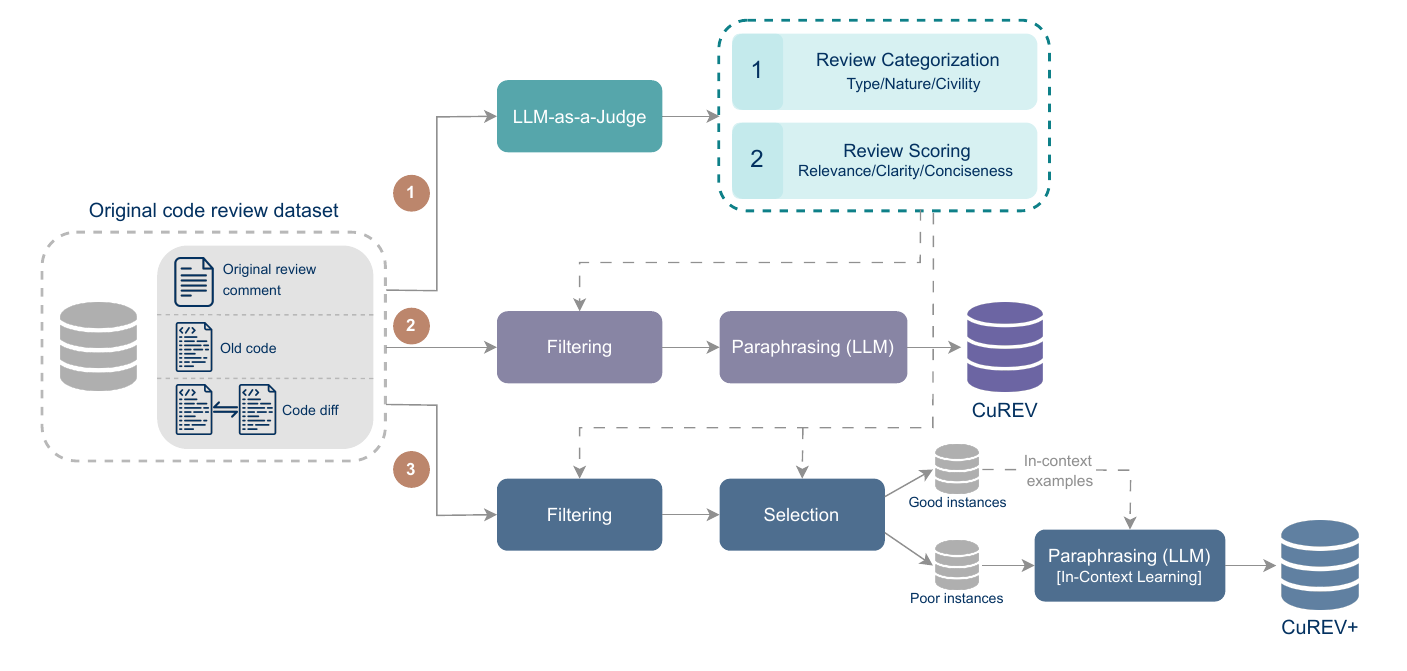}
    \caption{Overview of the two curation pipelines. \emph{CuREV} combines filtering and reformulation of poor-quality comments, while \emph{CuREV+} enhances this process by incorporating selection and in-context examples from high-quality reviews to guide reformulation.}
    \label{fig:approach}
\end{figure*}

In this section, we first recall the curation pipeline proposed in our previous work~\citep{sghaier2025harnessing}, which aimed to enhance the overall quality of large-scale code review datasets through systematic reformulation. Building on this foundation, we introduce an improved pipeline designed to enhance comment quality while preserving and enriching the diversity of writing styles. Both pipelines address common issues such as irrelevance, verbosity, incivility, and lack of clarity in review comments, ensuring that the resulting data is more reliable, informative, and useful for downstream tasks. The curated datasets produced through these pipelines, namely \emph{CuREV} and its enhanced version \emph{CuREV+}, are illustrated in \Fig{fig:approach}.

\subsection{Experimental Setup}

Based on our previous experimental findings, we developed two curation pipelines aimed at enhancing the overall quality of the selected code review dataset.

\paragraph*{\textbf{CuREV pipeline\\}}
The first curation pipeline, \emph{CuREV}, follows a two-step process: filtering and reformulation.  
In the filtering phase, we identify and remove irrelevant review comments, which cannot be improved through rephrasing. Relevance is considered an intrinsic property of a review: it either addresses a valid concern or it does not. Thus, unlike clarity or civility, it cannot be enhanced through linguistic reformulation.  
To apply this filtering, we rely on the relevance scores obtained from the initial evaluation (see \Sect{sec:initdata}) and set a threshold of $4$. Comments scoring below this threshold are removed, resulting in the elimination of $5{,}895$ samples and leaving $170{,}718$ relevant reviews.

These are three examples of review comments that were filtered out due to their low relevance scores ($relevance\_score<4$):

\begin{quote}
\textbf{Example 1:} \texttt{Need some edit here?}\\

\textbf{Example 2:} \texttt{Same here etc :)}\\

\textbf{Example 3:} \texttt{This is gross}
\end{quote}

Although such comments may make sense to human reviewers familiar with the context, since in real-world settings, much communication occurs verbally and directly, these comments remain irrelevant and unhelpful for automated models, as they provide no actionable information to learn from, and are therefore discarded.

In the second step, we focus on reformulating review comments to enhance the criteria of civility, clarity, and conciseness. For this task, we employed the LLM \emph{Llama-3.1-70B}, which was provided with specific instructions to guide the reformulation process. The reformulation task was designed to maintain the original intent and content of each comment while improving its form and presentation. \Table{tab:reform_prompt} presents an excerpt of the reformulation prompt, where these attributes are explicitly defined and reinforced.

\begin{table}[!t]
\centering
\caption{Excerpt of the prompt for comment reformulation.}
\label{tab:reform_prompt}
\begin{tabularx}{\linewidth}{X}
\toprule

\textbf{\#\#\# Review comment reformulation}

Your task is to reformulate and improve the given review comment by making it civil, clear, and concise without changing its core message or intent. 

\textsc{\textbf{1. Conciseness:}} Convey the message in the fewest words possible while remaining informative. Remove redundancy and irrelevant details.\\[0.3em]
\textsc{\textbf{2. Clarity:}} Ensure the comment is well-structured, grammatically correct, and unambiguous.\\[0.3em]
\textsc{\textbf{3. Civility:}} Maintain a respectful, professional, and constructive tone.\\[0.3em]
\textbf{\#\#\# Given review comment}\\
\{\textit{review\_comment}\}\\
\textbf{\#\#\# Code changes}\\
\{\textit{code\_diff}\}\\

\bottomrule
\end{tabularx}
\end{table}

The reformulation guidelines, employed in the prompt, emphasize three key criteria: conciseness, clarity, and civility. The reformulated comment should convey its message in the fewest words necessary while remaining informative, removing any redundancy or irrelevant details (\emph{conciseness}). It should also be well-structured, straightforward, and free of ambiguity to ensure ease of understanding (\emph{clarity}). Additionally, the comment must maintain a respectful and professional tone, providing constructive feedback without using harsh or inappropriate language (\emph{civility}).

The code review dataset used is exclusively composed of English comments, though some comments may include non-English words depending on the context. A powerful and multilingual LLM, such as Llama-3.1-70B, can appropriately handle these cases effectively. This was supported by findings in the literature, which indicate that LLMs, despite being predominantly trained on large amounts of English data, can still manage non-English languages to a reasonable extent~\citep{terryn2024exploratory}. 

In the final step, we re-evaluated the curated review comments using the same evaluation schema as was applied to the original dataset. 
Relevance was excluded from re-evaluation since reformulation cannot alter the underlying meaning or target of a comment; relevance is dependent on the content of the comment rather than its form.
This re-evaluation allowed for a direct comparison between the original and curated review comments, assessing variations in clarity, conciseness, and civility.

\paragraph*{\textbf{CuREV+ pipeline\\}}

The second pipeline, \emph{CuREV+}, follows the same main steps as \emph{CuREV} (\ie filtering and reformulation) but extends the process through a quality-based \emph{selection} mechanism and the use of \emph{in-context examples} to guide reformulation.

After filtering irrelevant comments (as in CuREV), the dataset is further partitioned into \emph{high-quality} and \emph{low-quality} comments according to civility, clarity, and conciseness scores. A comment is labeled as \emph{low-quality} if it is uncivil or if either its clarity or conciseness score falls below the threshold of $7$. In contrast, a comment is labeled as \emph{high-quality} if it is civil and obtains scores greater than or equal to $7$ in both clarity and conciseness. Following this procedure, the filtered dataset of $170{,}718$ comments consists of $77{,}824$ low-quality comments and $92{,}894$ high-quality comments.

The low-quality comments are selected for reformulation (\ie we need to enhance their quality through rephrasing), while the high-quality ones are retained unchanged. This decision ensures that we preserve the natural writing style and authenticity of human reviewers, avoiding over-normalization and loss of diversity of the dataset.  

For each low-quality comment, we randomly sample $10$ high-quality comments from the pool of preserved examples (\ie good comments). These are provided to the model as \emph{in-context examples} for inspiration during reformulation.  
The prompt used for CuREV+ is similar to that of CuREV but augmented with these high-quality reference comments, as shown in \Table{tab:reform_prompt2}.

\begin{table}[!t]
\centering
\caption{Excerpt of the prompt for comment reformulation with in-context examples.}
\label{tab:reform_prompt2}
\begin{tabularx}{\linewidth}{X}
\toprule
\textbf{\#\#\# Review comment reformulation with in-context examples}

Your task is to reformulate and improve the given review comment by making it civil, clear, and concise without changing its core message or intent. 
Use the provided high-quality examples as inspiration to guide the reformulation style and tone while ensuring consistency with the same quality attributes.

\textsc{\textbf{1. Conciseness:}} Convey the message in the fewest words possible while remaining informative. Remove redundancy and irrelevant details.\\[0.3em]
\textsc{\textbf{2. Clarity:}} Ensure the comment is well-structured, grammatically correct, and unambiguous.\\[0.3em]
\textsc{\textbf{3. Civility:}} Maintain a respectful, professional, and constructive tone.\\[0.3em]
\textbf{\#\#\# Examples of good review comments (for inspiration only)}\\
Inspire from the following list of good review comments.\\
\{\textit{example\_1}\}\\
\{\textit{example\_2}\}\\
\hspace{3em}\vdots\\
\{\textit{example\_10}\}\\
\textbf{\#\#\# Given review comment}\\
\{\textit{review\_comment}\}\\
\textbf{\#\#\# Code changes}\\
\{\textit{code\_diff}\}\\
\bottomrule
\end{tabularx}
\end{table}

This in-context setup allows the model to draw inspiration from human-written, high-quality examples, helping it internalize better stylistic and structural patterns during reformulation.  
Moreover, by varying the set of examples for each low-quality comment, we ensure that the model is exposed to a wide range of writing styles and tones. This randomized sampling promotes stylistic diversity and reduces the risk of producing overly uniform and artificial comments.  

The reformulation process thus remains guided by the same quality attributes (\ie clarity, conciseness, and civility) but is now enriched by exposure to exemplary human-authored comments. This approach enables the generation of more natural, contextually coherent, and human-like review comments while simultaneously preserving diversity and reducing homogenization.

\paragraph*{\textbf{Re-evaluation of curated datasets\\}}
Following both curation processes, we re-evaluate the resulting datasets (both of size $170,718$) using the same evaluation framework (\Sect{sec:initdata}) to ensure consistency and comparability. Since relevance is content-dependent, it is excluded from this re-assessment.  
The evaluation thus focuses on quantifying improvements in clarity, conciseness, and civility between the original dataset, \emph{CuREV}, and \emph{CuREV+}, which enables us to measure the effectiveness of each curation pipeline in producing higher-quality review comments.

%
%

\subsection{Results}

\begin{figure}[!t]
    \centering
    \begin{subfigure}{0.48\linewidth}
        \centering
        \includegraphics[width=\linewidth]{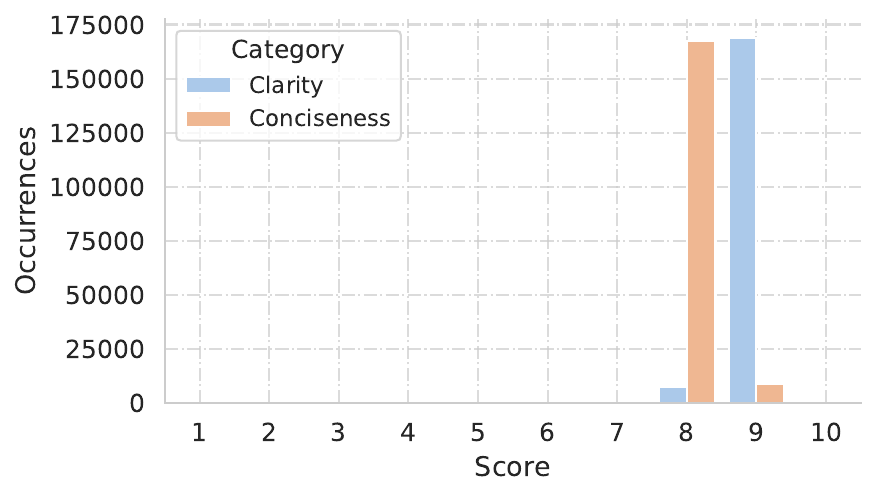}
        \caption{CuREV}
        \label{fig:curated_scoring_dist}
    \end{subfigure}
    \hfill
    \begin{subfigure}{0.48\linewidth}
        \centering
        \includegraphics[width=\linewidth]{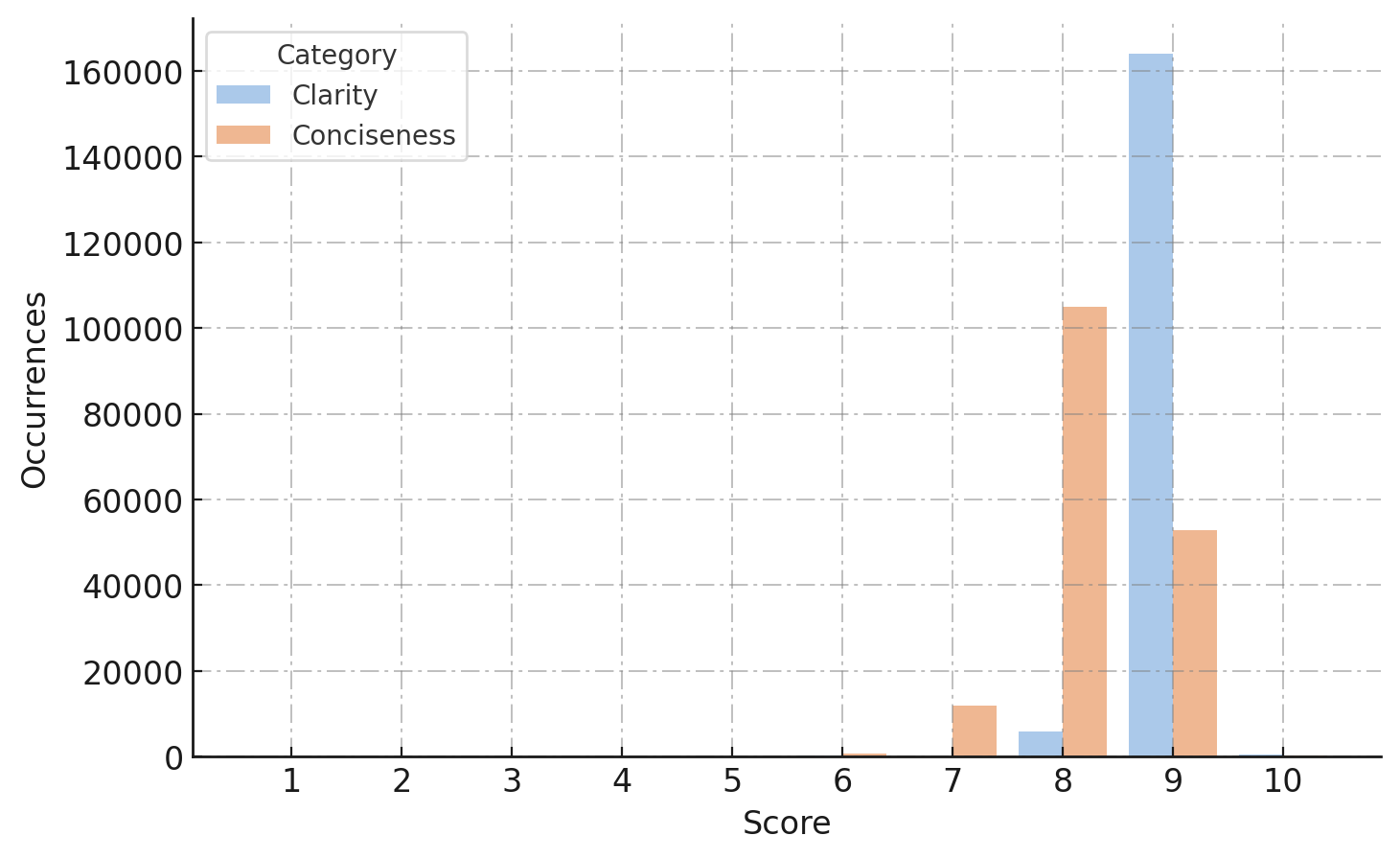}
        \caption{CuREV+}
        \label{fig:curated+_scoring_dist}
    \end{subfigure}
    \caption{Distribution of the clarity and conciseness scoring criteria across curated datasets.}
    \label{fig:curated_scoring_comparison}
\end{figure}

The analysis of the scoring criteria across the curated datasets, \ie CuREV and CuREV+, highlights the complementary benefits of the two proposed pipelines. As reported in \Table{tab:cur_scoring_dist} and \Fig{fig:curated_scoring_comparison}, both CuREV and CuREV+ substantially improve clarity and conciseness compared to the original dataset, albeit with different emphases.

\paragraph{\textbf{Clarity.}} 
Both curation strategies strongly enhanced clarity, with average scores rising from $6.89$ in the original dataset to $8.96$ in CuREV and $8.95$ in CuREV+. These gains (around $+30\%$ relative improvement) confirm that both curation pipelines systematically produce better-structured comments, reduce noise, and ensure the use of precise, grammatically correct language. 

Interestingly, both CuREV and CuREV+ achieve almost identical levels of clarity, although CuREV+ does not reformulate the entire dataset. In CuREV+, only $77{,}824$ comments were reformulated while $92{,}894$, which already have good quality, were left untouched. This selective strategy minimizes the number of altered or artificially rephrased comments, thereby preserving the naturalness and authenticity of human-written feedback. Yet, even with fewer interventions, CuREV+ reaches nearly the same clarity level as CuREV, demonstrating that clarity is a robust outcome of both curation pipelines and that selective reformulation can yield high-quality results without excessively compromising the human-like nature of the dataset.

\paragraph{\textbf{Conciseness.}} 
Improvements in conciseness show a sharper contrast between the two pipelines. While CuREV produced only modest gains ($7.71 \rightarrow 8.05$, $+4.4\%$), CuREV+ raised the average conciseness score to $8.53$ ($+10.6\%$). 

As shown in \Table{tab:cur_scoring_dist}, and \Fig{fig:curated_scoring_comparison}, CuREV+ significantly reduced verbosity compared to CuREV and the original dataset. This can be explained by its selective mechanism: it focuses reformulation efforts on ``poor'' comments while keeping ``good'' human-written comments (\ie concise and clear comments) intact, thus preserving natural brevity and reducing the tendency of LLMs to generate overly elaborate text.

\begin{table}[!t]
  \centering
  \caption{Evolution of the scoring criteria per category across the curated datasets (CuREV and CuREV+), expressed as percentage improvements over the original dataset.}
  \label{tab:cur_scoring_dist}
  \begin{tabular}{>{\centering\arraybackslash}p{1cm} >{\centering\arraybackslash}p{3cm} *{2}{c} *{2}{c}}
    \toprule
    \multirow{2}{*}{\textbf{Category}} & \multirow{2}{*}{\textbf{Subcategory}} 
    & \multicolumn{2}{c}{\textbf{CuREV}} & \multicolumn{2}{c}{\textbf{CuREV+}} \\
    \cmidrule(lr){3-4} \cmidrule(lr){5-6}
    & & \textbf{Clarity} & \textbf{Conciseness} & \textbf{Clarity} & \textbf{Conciseness} \\
    \midrule
    \multirow{6}{*}{\textbf{Type}} 
      & Refactoring   & $8.95$ (\up $14.9\%$) & $8.06$ (\up $15.3\%$) & $8.96$ (\up $15.0\%$) & $8.23$ (\up $17.7\%$) \\
      & Bugfix        & $8.98$ (\up $16.0\%$) & $8.03$ (\up $17.4\%$) & $8.96$ (\up $15.8\%$) & $8.91$ (\up $30.2\%$) \\
      & Testing       & $8.98$ (\up $13.4\%$) & $8.03$ (\up $15.2\%$) & $8.97$ (\up $13.3\%$) & $8.66$ (\up $24.3\%$) \\
      & Logging       & $8.97$ (\up $14.4\%$) & $8.02$ (\up $17.1\%$) & $8.97$ (\up $14.4\%$) & $8.18$ (\up $19.4\%$) \\
      & Documentation & $8.98$ (\up $17.8\%$) & $8.02$ (\up $19.3\%$) & $8.97$ (\up $17.7\%$) & $8.75$ (\up $30.2\%$) \\
      & Other         & $8.95$ (\up $30.5\%$) & $8.05$ (\up $36.4\%$) & $8.93$ (\up $30.2\%$) & $8.35$ (\up $41.5\%$) \\
    \midrule
    \multirow{4}{*}{\textbf{Nature}} 
      & Descriptive   & $8.76$ (\up $32.1\%$) & $8.02$ (\up $43.0\%$) & $8.80$ (\up $32.7\%$) & $8.21$ (\up $46.4\%$) \\
      & Prescriptive  & $8.96$ (\up $12.7\%$) & $8.05$ (\up $11.9\%$) & $8.96$ (\up $12.7\%$) & $8.57$ (\up $19.2\%$) \\
      & Clarification & $8.96$ (\up $18.3\%$) & $8.03$ (\up $20.6\%$) & $8.94$ (\up $18.1\%$) & $8.39$ (\up $26.0\%$) \\
      & Other         & $9.00$ (\up $104.5\%$) & $8.00$ (\up $94.2\%$) & $8.25$ (\up $87.5\%$) & $8.50$ (\up $106.3\%$) \\
    \midrule
    \multirow{2}{*}{\textbf{Civility}} 
      & Civil   & $8.96$ (\up $15.6\%$) & $8.05$ (\up $16.2\%$) & $8.95$ (\up $15.6\%$) & $8.53$ (\up $23.1\%$) \\
      & Uncivil & -- & -- & -- & -- \\
    \midrule
    \multirow{1}{*}{\textbf{Average}} 
      & -- & $8.96$ (\up $30.0\%$) & $8.05$ (\up $4.4\%$) & $8.95$ (\up $29.9\%$) & $8.53$ (\up $10.6\%$) \\
    \bottomrule
  \end{tabular}
\end{table}

\Table{tab:cur_categories_dist} presents the evolution of comment categorizations in the curated dataset compared to the original one. In the \emph{nature} category, the percentage of \emph{prescriptive} comments increased to $90.20\%$ (vs. $62.6\%$ in the original dataset) and the percentage of descriptive comments decreased significantly to $0.95\%$, demonstrating a strong shift towards more directive, explicit, and actionable guidance to developers. This could be explained by the fact that prescriptive comments might be clearer compared to descriptive comments.

In the \emph{civility} category, all comments were marked as \emph{civil}, achieving a perfect score ($100\%$) and eliminating any \emph{Uncivil} comments. This indicates that the curation process effectively addressed concerns related to inappropriate or harsh language, creating a more professional and constructive dataset.

\begin{table}[!htbp]
  \centering
  \caption{Categories statistics in the curated datasets (CuREV and CuREV+).}
  \label{tab:cur_categories_dist}
  \begin{tabular}{>{\centering\arraybackslash}p{1.2cm} >{\centering\arraybackslash}p{3cm} *{2}{>{\centering\arraybackslash}p{2cm} >{\centering\arraybackslash}p{1.8cm}} }
    \toprule
    \multirow{2}{*}{\textbf{Category}} & \multirow{2}{*}{\textbf{Subcategory}} 
    & \multicolumn{2}{c}{\textbf{CuREV}} 
    & \multicolumn{2}{c}{\textbf{CuREV+}} \\
    \cmidrule(lr){3-4} \cmidrule(lr){5-6}
    & & \textbf{Count} & \textbf{Percentage} & \textbf{Count} & \textbf{Percentage} \\
    \midrule
    \multirow{4}{*}{\textbf{Nature}} 
      & Descriptive   & $1,621$   & $0.95$\down  & $3,255$   & $1.91$\down \\
      & Prescriptive  & $153,987$ & $90.20$\up   & $157,376$ & $92.18$\up \\
      & Clarification & $28,595$  & $16.75$\down & $33,212$  & $19.45$\down \\
      & Other         & $16,901$  & $9.90$\up    & $4$       & $0.00$\down \\
    \midrule
    \multirow{2}{*}{\textbf{Civility}} 
      & Civil   & $170,718$ & $100$\up & $170,718$ & $100$\up \\
      & Uncivil & $0$       & $0$\down & $0$       & $0$\down \\
    \bottomrule
  \end{tabular}
\end{table}

\paragraph{\textbf{Nature and Conciseness.}} 
Shifts in the distribution of categories, presented in \Table{tab:cur_categories_dist}, further illustrate differences between the two curated datasets compared to the original one. In the \emph{nature} category, CuREV strongly increased the proportion of prescriptive comments ($62.6\% \rightarrow 90.2\%$), while CuREV+ pushed this trend even further ($92.2\%$), reflecting a dataset that overwhelmingly favors actionable and directive feedback. Clarification comments also rose under CuREV+, indicating that the second pipeline helped surface questions and justifications alongside prescriptive comments. 

In the \emph{civility} category, the original dataset contains uncivil reviews with inappropriate language and a lack of professionalism, which could impede effective communication and contribute to an unconstructive code review environment. Both pipelines achieved the complete elimination of uncivil comments, yielding a fully civil dataset.

\paragraph{\textbf{Comparative Insights.}} 
Overall, CuREV can be seen as a broad reformulation strategy that homogenizes the dataset, boosting clarity and prescriptiveness at scale. CuREV+, in contrast, preserves naturally good comments while selectively refining poor ones. This selective approach leads to better conciseness and a more balanced distribution of review styles, mitigating the risk of oversynthetic, uniform, or ``LLM-like'' phrasing.

\paragraph{\textbf{Examples of Review Comments from Original, CuREV, and CuREV+ datasets.}}

The results obtained, presented in \Table{tab:reform_examples}, present examples of review comments drawn from the original dataset, CuREV, and CuREV+. These examples allow us to directly compare the two curation pipelines and illustrate how they operate differently, leading to distinct results.  



\begin{table*}[!htbp]
\centering
\caption{Examples of review comments across the original dataset, CuREV, and CuREV+.}
\label{tab:reform_examples}
\begin{tabularx}{\textwidth}{lX}
\toprule
\multicolumn{1}{c}{\textbf{Version}} & \multicolumn{1}{c}{\textbf{Review Comment}} \\
\midrule

\rowcolor{gray!20}\multicolumn{2}{c}{\textbf{Example 1}}\\\midrule

\textbf{Original} & 
\texttt{mmmm welcome to the fucking world of TS ids... -1 or 0 for invalid?} \\ \midrule

\textbf{CuREV} \newline{\scriptsize [Clarity $\uparrow$, Civility $\uparrow$]} & 
\texttt{Consider returning 0 instead of -1 for invalid object IDs to maintain consistency with standard error handling conventions.} \\ \midrule

\textbf{CuREV+} \newline{\scriptsize [Clarity $\uparrow$, Civility $\uparrow$, Conciseness $\uparrow$]} & 
\texttt{Instead of using -1 or 0 to indicate an invalid result, can we use a more explicit value or an exception to convey the result of \`findItemByObjectId\`? This would improve the method's return value to better indicate the presence or absence of an item.} \\ \midrule

\rowcolor{gray!20}\multicolumn{2}{c}{\textbf{Example 2}}\\\midrule

\textbf{Original} & 
\texttt{why is the modifier "package", better to change to private} \\ \midrule

\textbf{CuREV} \newline{\scriptsize [Clarity $\uparrow$]} & 
\texttt{Consider changing the access modifier of `booleanProperties` from "package" to "private" for consistency and encapsulation.} \\ \midrule

\textbf{CuREV+} \newline{\scriptsize [Clarity $\uparrow$, Conciseness $\uparrow$]} & 
\texttt{The modifier for `booleanProperties` should be `private` for better encapsulation, not `package`} \\ \midrule

\rowcolor{gray!20}\multicolumn{2}{c}{\textbf{Example 3}}\\\midrule

\textbf{Original} & 
\texttt{Let's not add this field for datasets which do not have release note yet.} \\ \midrule

\textbf{CuREV} \newline{\scriptsize [Clarity $\uparrow$]} & 
\texttt{Consider excluding the RELEASE\_NOTES field for datasets without release notes to avoid unnecessary additions.} \\ \midrule

\textbf{CuREV+} \newline{\scriptsize [Clarity $\uparrow$, Conciseness $\uparrow$]} & 
\texttt{Can we avoid adding the 'RELEASE\_NOTES' field for datasets without release notes yet?} \\ \midrule

\rowcolor{gray!20}\multicolumn{2}{c}{\textbf{Example 4}}\\\midrule

\textbf{Original} & 
\texttt{`type ReactionData` maybe?} \\ \midrule

\textbf{CuREV} \newline{\scriptsize [Clarity $\uparrow$]} & 
\texttt{Consider defining a type for the reaction data, e.g., `ReactionData`.} \\ \midrule

\textbf{CuREV+} \newline{\scriptsize [Clarity $\uparrow$, Conciseness $\uparrow$]} & 
\texttt{Could we explicitly type the ReactionData as a specific type, for example, ReactionData: FooType, to improve code clarity and type safety?} \\ \midrule

\rowcolor{gray!20}\multicolumn{2}{c}{\textbf{Example 5}}\\\midrule

\textbf{Original} & 
\texttt{Move constants at the top of the class, where they should be} \\ \midrule

\textbf{CuREV} \newline{\scriptsize [Clarity $\uparrow$, Conciseness $\uparrow$]} & 
\texttt{Consider placing constants at the top of the class for better organization and adherence to conventional coding practices.} \\ \midrule

\textbf{CuREV+} & 
\emph{Not reformulated (original retained)} \\ \midrule

\rowcolor{gray!20}\multicolumn{2}{c}{\textbf{Example 6}}\\\midrule

\textbf{Original} & 
\texttt{@username can you please verify this change again? Looks incorrect.} \\ \midrule

\textbf{CuREV} \newline{\scriptsize [Clarity $\uparrow$]} & 
\texttt{Could you please re-verify the change to pass `null` as the `cancelChecker` argument in the `completion` method? I'm concerned it might be incorrect.} \\ \midrule

\textbf{CuREV+} & 
\emph{Not reformulated (original retained)} \\ \midrule

\rowcolor{gray!20}\multicolumn{2}{c}{\textbf{Example 7}}\\\midrule

\textbf{Original} & 
\texttt{site may not be always initiated when this is called. So important to check `if getattr(frappe.local, "site", None):`} \\ \midrule

\textbf{CuREV} \newline{\scriptsize [Clarity $\uparrow$]} & 
\texttt{Consider adding a null check for `frappe.local.site` to ensure it's initialized before accessing it, as it may not always be set when this function is called.} \\ \midrule

\textbf{CuREV+} & 
\emph{Not reformulated (original retained)} \\ \midrule

\rowcolor{gray!20}\multicolumn{2}{c}{\textbf{Example 8}}\\\midrule

\textbf{Original} & 
\texttt{We can remove the SERVER\_CONNECTOR\_TEST\_PORT and move it to the Listener} \\ \midrule

\textbf{CuREV} \newline{\scriptsize [Clarity $\uparrow$]} & 
\texttt{Consider moving the constant `SERVER\_CONNECTOR\_TEST\_PORT` to the Listener class, as it seems more relevant to its functionality.} \\ \midrule

\textbf{CuREV+} & 
\emph{Not reformulated (original retained)} \\ \midrule

\rowcolor{gray!20}\multicolumn{2}{c}{\textbf{Example 9}}\\\midrule

\textbf{Original} & 
\texttt{Why was this change made?} \\ \midrule

\textbf{CuREV} \newline{\scriptsize [Clarity $\uparrow$]} & 
\texttt{Can you provide the reasoning behind adding the `WriteTo` call before closing the writer? This change seems to introduce a new dependency, and I'd like to understand the motivation behind it.} \\ \midrule

\textbf{CuREV+} & 
\emph{Not reformulated (original retained)} \\ \midrule

\rowcolor{gray!20}\multicolumn{2}{c}{\textbf{Example 10}}\\\midrule

\textbf{Original} & 
\texttt{If this is a prescribed Django fix for this, perhaps link to the permalink in docs here to explain what this is for.} \\ \midrule

\textbf{CuREV} \newline{\scriptsize [Clarity $\uparrow$, Conciseness $\uparrow$]} & 
\texttt{Consider adding a link to the relevant Django documentation to explain the purpose of this fix, as it appears to be a prescribed solution.} \\ \midrule

\textbf{CuREV+} &  
\emph{Not reformulated (original retained)} \\

\bottomrule
\end{tabularx}
\end{table*}


CuREV systematically reformulated \emph{all} comments in a uniform way, resulting in highly consistent phrasing (Examples~1--10 all start with ``Consider''). In contrast, CuREV+ introduced natural variation by reformulating only problematic comments while preserving high-quality ones in their original form. As a result, CuREV+ produces comments that are both clear and diverse in style, with openings such as ``Instead of'', ``The modifier'', ``Can we'', or ``Could we'', which feel more authentic and human-like.  

Examples~1--4 highlight cases where the original comments lacked conciseness, clarity, or civility. In these cases, CuREV reformulates them into concise and polite directives, while CuREV+ goes further by offering stylistically varied reformulations through the use of in-context examples for inspiration. For instance, the uncivil phrasing in Example~1 (\texttt{mmmm welcome to the fucking world of TS ids...}) is rephrased into professional and instructive feedback by both pipelines. Similar patterns appear in Examples~2--4, where vague or incomplete comments are transformed into more precise and context-aware guidance and where CuREV+ introduces additional variety, making the suggestion feel less formulaic, compared to CuREV.  

In contrast, Examples~5--10 represent comments already deemed high-quality—clear, civil, and concise. These are preserved in CuREV+ exactly as in the original dataset and thus excluded from reformulation. For instance, Example~5 (\texttt{Move constants at the top of the class, where they should be}) already provides a precise and polite suggestion, requiring no further editing. CuREV, however, reformulates even these high-quality comments, which may reduce the naturalness and stylistic diversity of the dataset.

In summary, CuREV emphasizes consistency and quality by reformulating every comment into a uniform style, while CuREV+ strikes a balance between quality and diversity. By selectively targeting only low-quality comments and retaining natural high-quality ones, CuREV+ achieves both improved quality and richer stylistic diversity. This makes CuREV+ particularly valuable for training models that must generate feedback that is not only clear, concise, and civil, but also representative of the authentic variation found in real-world code reviews.

\begin{center}
\begin{tcolorbox}[
    colback=purple!5!white,
    colframe=purple!75!black,
    left=3mm,
    overlay={\draw[purple!75!black, line width=3pt] (frame.south west)--(frame.north west);},
    title=\textbf{Answer to RQ2 (Quality of Curated Datasets)},
    fonttitle=\bfseries,
    boxrule=0pt,
    arc=2pt,
    before skip=5pt,
    after skip=5pt
]
The results show that both curation pipelines markedly improve review comments compared to the original dataset. CuREV yielded clearer ($6.89 \rightarrow 8.96$), more prescriptive ($62.6\% \rightarrow 90.2\%$), more concise ($7.71 \rightarrow 8.05$), and more civil ($98.8\% \rightarrow 100\%$) comments. CuREV+ achieved similar clarity improvements ($6.89 \rightarrow 8.95$), but more substantial gains in conciseness ($7.71 \rightarrow 8.53$), while reinforcing prescriptiveness ($62.6\% \rightarrow 92.2\%$) and sustaining perfect civility.
\end{tcolorbox}
\end{center}

\subsection{Semantic Preservation and BERTScore Analysis}
To formally measure semantic preservation between the original and reformulated comments beyond standard lexical n-gram limitations, we integrated \textbf{BERTScore} \citep{zhang2019bertscore} as a foundational embedding-based evaluation metric utilizing contextual RoBERTa-large embeddings. 

Our computations yield corroborating quantitative evidence for our selective curation technique. For the CuREV+ pipeline processing defective reviews ($N=77,824$), it achieved a strong BERTScore F1 of $0.896 \pm 0.034$. In contrast, CuREV's uniform application across the full $N=170,718$ dataset averaged $0.836 \pm 0.019$. These scores demonstrate that selectively reformulating only poor-quality feedback effectively preserves the underlying semantic and technical intent significantly better than monolithic, uniform generation.

\subsection{Sensitivity Analysis of Quality Thresholds}
To validate our selection of curation thresholds (Relevance=4, Quality=6), we conducted a comprehensive sensitivity analysis detailing the impact of varying the thresholds on the overall composition of our dataset.

\begin{table}[!t]

\centering
\caption{Sensitivity analysis of threshold parameters on dataset partitioning.}
\label{tab:sensitivity}
\begin{tabular}{cc|rrr}
\toprule
\textbf{Relevance} & \textbf{Quality} & \textbf{Irrelevant} & \textbf{Low-quality} & \textbf{High-quality} \\
\textbf{Threshold} & \textbf{Threshold} & & & \\
\midrule
$2$ & $4$ & 1,511 (0.86\%) & 13,100 (7.48\%) & 162,002 (92.52\%) \\
$2$ & $5$ & 1,511 (0.86\%) & 13,250 (7.57\%) & 161,852 (92.43\%) \\
$2$ & $6$ & 1,511 (0.86\%) & 82,186 (46.94\%) & 92,916 (53.06\%) \\
$2$ & $7$ & 1,511 (0.86\%) & 127,783 (72.98\%) & 47,319 (27.02\%) \\
$2$ & $8$ & 1,511 (0.86\%) & 152,655 (87.18\%) & 22,447 (12.82\%) \\
$3$ & $4$ & 4,834 (2.74\%) & 10,212 (5.94\%) & 161,567 (94.06\%) \\
$3$ & $5$ & 4,834 (2.74\%) & 10,359 (6.03\%) & 161,420 (93.97\%) \\
$3$ & $6$ & 4,834 (2.74\%) & 78,885 (45.92\%) & 92,894 (54.08\%) \\
$3$ & $7$ & 4,834 (2.74\%) & 124,482 (72.47\%) & 47,297 (27.53\%) \\
$3$ & $8$ & 4,834 (2.74\%) & 149,332 (86.93\%) & 22,447 (13.07\%) \\
$4$ & $4$ & 5,895 (3.34\%) & 9,175 (5.37\%) & 161,543 (94.63\%) \\
$4$ & $5$ & 5,895 (3.34\%) & 9,321 (5.46\%) & 161,397 (94.54\%) \\
$4$ & $6$ & 5,895 (3.34\%) & 77,824 (45.59\%) & 92,894 (54.41\%) $\star$ \\
$4$ & $7$ & 5,895 (3.34\%) & 123,421 (72.3\%) & 47,297 (27.7\%) \\
$4$ & $8$ & 5,895 (3.34\%) & 148,271 (86.85\%) & 22,447 (13.15\%) \\
$5$ & $4$ & 6,955 (3.94\%) & 8,954 (5.28\%) & 160,704 (94.72\%) \\
$5$ & $5$ & 6,955 (3.94\%) & 9,098 (5.36\%) & 160,560 (94.64\%) \\
$5$ & $6$ & 6,955 (3.94\%) & 76,764 (45.25\%) & 92,894 (54.75\%) \\
$5$ & $7$ & 6,955 (3.94\%) & 122,361 (72.12\%) & 47,297 (27.88\%) \\
$5$ & $8$ & 6,955 (3.94\%) & 147,211 (86.77\%) & 22,447 (13.23\%) \\
$6$ & $4$ & 7,020 (3.97\%) & 8,924 (5.26\%) & 160,669 (94.74\%) \\
$6$ & $5$ & 7,020 (3.97\%) & 9,063 (5.34\%) & 160,530 (94.66\%) \\
$6$ & $6$ & 7,020 (3.97\%) & 76,699 (45.23\%) & 92,894 (54.77\%) \\
$6$ & $7$ & 7,020 (3.97\%) & 122,296 (72.11\%) & 47,297 (27.89\%) \\
$6$ & $8$ & 7,020 (3.97\%) & 147,146 (86.76\%) & 22,447 (13.24\%) \\
\bottomrule
\end{tabular}

\vspace{0.5em}
\raggedright
\footnotesize{$\star$ denotes the configuration used in our study.}
\end{table}

The analysis confirms that our chosen thresholds represent a principled balance point, safely filtering out entirely irrelevant feedback while preserving a statistically robust 54.4\% of high-quality reviews without over-purging.

\section{Diversity of curated datasets}
\label{sec:diversity}

To better understand the linguistic variety and stylistic tendencies of the curated datasets, we measure and compare the \textbf{diversity} of both CuREV and CuREV+. Both datasets have the same size ($170,718$), ensuring that differences observed are due to curation strategies rather than scale.

\subsection{Diversity Metrics.}
To assess the lexical and structural variety of review comments in CuREV and CuREV+, we employ several established metrics. 
We first compute \textbf{unigram, bigram, and trigram diversity}, defined as the ratio of distinct $n$-grams over the total number of $n$-grams in the dataset:
\[
\text{Diversity}_n = \frac{\text{\# distinct $n$-grams}}{\text{\# total $n$-grams}}, \quad n \in \{1,2,3\}.
\]
This captures the breadth of vocabulary at the token level as well as the richness of multiword expressions~\citep{zhou2021evaluating,li2015diversity}. 
We also measure \textbf{Self-BLEU}~\citep{zhu2018texygen}, which quantifies redundancy by averaging BLEU scores~\citep{papineni2002bleu} between each comment and the rest of the dataset. 
Lower Self-BLEU values indicate greater diversity, as comments are less repetitive. 
Finally, we compute the \textbf{entropy} of the unigram distribution,
\[
H = - \sum_{w \in V} p(w) \log p(w),
\]
where $V$ is the vocabulary and $p(w)$ is the empirical probability of token $w$. 
Higher entropy indicates a more uniform token distribution and thus greater lexical variability. 
Together, these complementary metrics provide a comprehensive view of diversity, balancing vocabulary breadth, phrase variety, and repetitiveness.

\paragraph{\textbf{Results.}}
\Table{tab:diversity_metrics} shows the values of the different diversity metrics for both CuREV and CuREV+.

\begin{table}[!htbp]
  \centering
  \caption{Diversity metrics for CuREV and CuREV+. Higher values for diversity metrics (unigram, bigram, trigram, entropy) indicate greater variety, while lower Self-BLEU indicates less repetition.}
  \label{tab:diversity_metrics}
  \begin{tabular}{lcc}
    \toprule
    \textbf{Metric} & \textbf{CuREV} & \textbf{CuREV+} \\
    \midrule
    Unigram diversity       & $0.0267$ & $0.0275$ \\
    Bigram diversity        & $0.2092$ & $0.2192$ \\
    Trigram diversity       & $0.4889$ & $0.5314$ \\
    Entropy                 & $6.6295$ & $6.6671$ \\
    Self-BLEU               & $0.2466$ & $0.1431$ \\
    \bottomrule
  \end{tabular}
\end{table}

Unigram diversity is nearly identical across the two datasets ($0.0267$ for CuREV vs. $0.0275$ for CuREV+), indicating that both preserve a comparable vocabulary breadth.  \textbf{Bigram and trigram diversity}, however, increase in CuREV+ ($0.2192$ vs. $0.2092$, and $0.5314$ vs. $0.4889$), showing that CuREV+ exhibits richer multi-word combinations and a broader repertoire of phrasing.

\textbf{Entropy} also rises slightly in CuREV+ ($6.67$ vs. $6.63$), pointing to a more uniform word distribution and reduced over-reliance on a limited subset of terms.  
Finally, \textbf{Self-BLEU}, a measure of internal repetitiveness, drops significantly in CuREV+ ($0.1431$ vs. $0.2466$). This suggests that CuREV+ reduces redundancy and avoids repeating the same formulations across comments.

In short, while CuREV and CuREV+ achieve similar vocabulary coverage, CuREV+ produces more diverse phrasings and less redundancy, which is important for training models that generalize well to unseen review contexts.

\subsection{$N$-gram distributions}

To gain a finer view of stylistic writing variety, we analyze the top $n$-grams (\Table{tab:ngrams_curev_vs_curevplus}).

\begin{table*}[!htbp]
\centering
\caption{Top $n$-grams (unigrams, bigrams, trigrams) for CuREV and CuREV+.}
\label{tab:ngrams_curev_vs_curevplus}
\begin{tabular}{clrclrclr}
\toprule
& \multicolumn{2}{c}{\textbf{Unigrams}} & \multicolumn{2}{c}{\textbf{Bigrams}} & \multicolumn{2}{c}{\textbf{Trigrams}} \\
\cmidrule(lr){2-3} \cmidrule(lr){4-5} \cmidrule(lr){6-7}
\textbf{Rank} & \textbf{Term} & \# & \textbf{Term} & \# & \textbf{Term} & \# \\
\midrule
\multicolumn{7}{c}{\textbf{CuREV}} \\
\midrule
1 & consider & 142{,}181 & consider adding & 28{,}338 & consider adding a & 19{,}111 \\
2 & using & 35{,}870 & consider using & 26{,}147 & consider removing the & 9{,}186 \\
3 & adding & 30{,}949 & instead of & 24{,}808 & consider using a & 6{,}932 \\
4 & instead & 27{,}984 & adding a & 19{,}856 & consider renaming the & 4{,}850 \\
5 & code & 23{,}077 & to ensure & 16{,}693 & consider simplifying the & 4{,}390 \\
6 & consistency & 21{,}509 & consider removing & 13{,}407 & the current implementation & 4{,}367 \\
7 & method & 20{,}439 & removing the & 10{,}986 & the reasoning behind & 3{,}950 \\
8 & ensure & 19{,}350 & for consistency & 10{,}877 & consider using the & 3{,}927 \\
9 & removing & 16{,}155 & to improve & 10{,}489 & you explain the & 3{,}618 \\
10 & error & 15{,}200 & consider renaming & 10{,}235 & to maintain consistency & 3{,}526 \\
\midrule
\multicolumn{7}{c}{\textbf{CuREV+}} \\
\midrule
1 & code & 41{,}541 & should be & 16{,}929 & to improve code & 4{,}485 \\
2 & should & 41{,}231 & it s & 16{,}462 & for better code & 4{,}283 \\
3 & use & 32{,}213 & the code & 15{,}822 & it s not & 4{,}130 \\
4 & instead & 19{,}656 & instead of & 15{,}642 & in the code & 3{,}974 \\
5 & better & 19{,}656 & can we & 12{,}143 & it would be & 3{,}611 \\
6 & here & 19{,}485 & i think & 7{,}539 & is there a & 2{,}927 \\
7 & using & 18{,}326 & can you & 6{,}730 & a more specific & 2{,}903 \\
8 & could & 17{,}706 & don t & 6{,}563 & more concise and & 2{,}612 \\
9 & specific & 16{,}968 & improve code & 6{,}174 & can we use & 2{,}304 \\
10 & method & 15{,}556 & should we & 5{,}889 & i don t & 2{,}304 \\
\bottomrule
\end{tabular}
\end{table*}

\paragraph{\textbf{CuREV.}}
CuREV is heavily dominated by repetitive functional constructions.

The most frequent unigram is ``\emph{consider}'' (142k occurrences), followed by tokens like ``\emph{using}'', ``\emph{adding}'', ``\emph{instead}'', or ``\emph{consistency}''.  
The top bigrams are similarly repetitive: ``\emph{consider adding}'' (28k), ``\emph{consider using}'' (26k), ``\emph{instead of}'' (25k).  
Trigrams emphasize this uniformity: ``\emph{consider adding a}'' (19k) and ``\emph{consider removing the}'' (9k).

This highlights the tendency of CuREV toward a single directive style centered on the word ``\emph{consider}''. Such uniformity may stem from the bias of the employed LLM to regard ``consider'' as a formal and professional entry point for review comments, aligning with a civil and non-confrontational tone when addressing developers.

\paragraph{\textbf{CuREV+.}}
By contrast, CuREV+ presents a more diverse lexical profile.

CuREV+ shows far fewer repeated unigrams, bigrams, and trigrams, suggesting a richer and more varied writing style. The top unigrams in CuREV+ are much more diverse (``\emph{code}'' 41k, ``\emph{should}'' 41k, ``\emph{use}'' 32k, ``\emph{better}'' 19k, ``\emph{specific}'' 17k), in contrast to the overwhelming dominance of ``\emph{consider}'' (142k) in CuREV. Similarly, bigrams and trigrams in CuREV+ emphasize more varied  phrasing, such as ``\emph{should be}'' (16k), ``\emph{it’s not}'' (4.1k), ``\emph{for better code}'' (4.3k), and ``\emph{can we use}'' (2.3k), while CuREV being dominated by formulaic sequences like ``\emph{consider adding}'' (28k) or ``\emph{consider using}'' (26k). This contrast illustrates that CuREV+ enables a richer and more diverse writing style, whereas CuREV tends toward a prescriptive monotony.

In addition, while CuREV is formulaic in structure, CuREV+ introduces more variety in both vocabulary and phrase construction, reflecting comments that are both directive (\emph{should be}, \emph{use a}) and reflective (\emph{can we use}, \emph{I don’t}).

\subsection{Prefix $n$-grams and writing style}

While standard $n$-gram statistics capture the overall lexical diversity of the datasets, they do not necessarily reflect how comments \emph{begin}, which often encodes tone, formality, and writing style. Prefix n-grams analysis addresses this by focusing exclusively on the leading tokens and phrases of comments, thereby highlighting recurrent openings and stylistic conventions. 

For instance, prefix analysis reveals even more pronounced stylistic contrasts between the two datasets (\Table{tab:prefix_ngrams_curev_vs_curevplus}). In CuREV, comments overwhelmingly start with the directive ``\emph{consider}'', reflecting a uniform prescriptive tone. In contrast, CuREV+ exhibits a wider range of prefixes, such as ``\emph{can we}'', ``\emph{should we}'', or ``\emph{please}'', pointing to a more conversational, collaborative, and varied reviewing style.

\begin{table*}[!htbp]
\centering
\caption{Top \emph{prefix} $n$-grams (unigrams, bigrams, trigrams) for CuREV and CuREV+.}
\label{tab:prefix_ngrams_curev_vs_curevplus}
\begin{tabular}{clrclrclr}
\toprule
& \multicolumn{2}{c}{\textbf{Unigrams}} & \multicolumn{2}{c}{\textbf{Bigrams}} & \multicolumn{2}{c}{\textbf{Trigrams}} \\
\cmidrule(lr){2-3} \cmidrule(lr){4-5} \cmidrule(lr){6-7}
\textbf{Rank} & \textbf{Term} & \# & \textbf{Term} & \# & \textbf{Term} & \# \\
\midrule
\multicolumn{7}{c}{\textbf{CuREV}} \\
\midrule
1 & consider & 126{,}340 & consider adding & 25{,}842 & consider adding a & 17{,}560 \\
2 & the & 14{,}434 & consider using & 24{,}094 & consider removing the & 8{,}576 \\
3 & can & 4{,}889 & consider removing & 10{,}705 & consider using a & 6{,}361 \\
4 & this & 4{,}272 & consider renaming & 9{,}571 & consider renaming the & 4{,}659 \\
5 & to & 3{,}470 & can you & 4{,}683 & consider simplifying the & 4{,}260 \\
6 & is & 2{,}867 & consider simplifying & 4{,}345 & consider using the & 3{,}495 \\
7 & instead & 1{,}883 & consider moving & 3{,}299 & consider moving the & 2{,}578 \\
8 & when & 1{,}439 & consider extracting & 2{,}646 & can you explain & 2{,}237 \\
9 & for & 1{,}242 & consider rephrasing & 2{,}414 & consider rephrasing the & 1{,}867 \\
10 & could & 1{,}035 & consider making & 2{,}388 & consider revising the & 1{,}672 \\
\midrule
\multicolumn{7}{c}{\textbf{CuREV+}} \\
\midrule
1 & can & 13{,}862 & can we & 8{,}514 & can we use & 1{,}734 \\
2 & consider & 10{,}797 & can you & 4{,}171 & consider using a & 1{,}620 \\
3 & suggestion & 7{,}932 & consider using & 4{,}071 & i don t & 1{,}088 \\
4 & please & 5{,}312 & should we & 3{,}628 & would it be & 909 \\
5 & could & 4{,}761 & i think & 3{,}563 & i think this & 782 \\
6 & why & 4{,}325 & could you & 2{,}368 & can you add & 703 \\
7 & use & 3{,}777 & let s & 2{,}028 & i think we & 687 \\
8 & nit & 2{,}526 & note that & 1{,}900 & can we add & 685 \\
9 & note & 2{,}089 & consider adding & 847 & consider adding a & 547 \\
10 & let & 2{,}045 & consider making & 825 & can you please & 512 \\
\bottomrule
\end{tabular}
\end{table*}

CuREV is overwhelmingly dominated by \emph{consider} as the first token: \emph{consider adding} ($26k$), \emph{consider using} ($24k$), \emph{consider removing} ($11k$). The unigram \emph{consider} alone accounts for $126k$ prefixes. This shows that CuREV enforces a uniform, almost templated reviewing style, where suggestions are nearly always framed as ``Consider \ldots''.  

CuREV+, however, distributes its prefixes across a wider set of pragmatic openings. While \emph{consider} remains common ($10k$), prefixes such as \emph{can we} ($8.5k$), \emph{should we} ($3.6k$), \emph{i think} ($3.6k$), \emph{please} ($5.3k$), \emph{why} ($4.3k$), and \emph{note that} ($1.9k$) reflect multiple interaction styles: collaborative questioning (\emph{can we \ldots}), prescriptive directives (\emph{should we \ldots}), politeness markers (\emph{please \ldots}), and explanatory tones (\emph{note that \ldots}).

This shift suggests that CuREV+ better preserves the natural heterogeneity of real reviewer-developer communication rather than all being homogenized into a single ``Consider \ldots'' template. This difference is particularly important for training downstream LLMs: CuREV may bias models toward rigid, repetitive outputs, while CuREV+ encourages more flexible and human-like review comment generation.

\begin{center}
\begin{tcolorbox}[
    colback=brown!5!white,
    colframe=brown!75!black,
    left=3mm,
    overlay={\draw[brown!80!black, line width=3pt] (frame.south west)--(frame.north west);},
    title=\textbf{Answer to RQ3 (Diversity)},
    fonttitle=\bfseries,
    boxrule=0pt,
    arc=2pt,
    before skip=5pt,
    after skip=5pt
]
CuREV exhibits a highly uniform style, with prefixes dominated by the directive ``\emph{consider}'', which the employed LLM may interpret as a formal and professional tone to address developers civilly. 
In contrast, CuREV+ demonstrates much less repetition of unigrams, bigrams, and trigrams, suggesting a richer and more varied writing style. 
Prefixes in CuREV+ include different forms such as ``\emph{can we}'', ``\emph{should we}'', and ``\emph{please}'', which indicate greater stylistic diversity. 
\end{tcolorbox}
\end{center}

\section{A Comparative Study of Initial and Curated Code Review Datasets on Downstream Code Review Tasks}
\label{sec:analysis}

\subsection{Impact of Curated Reviews on Comment Generation}
\label{subsec:model_data}
In this section, we investigate the impact of curated reviews on automating the comment generation process. By comparing models trained on the original dataset, the curated CuREV dataset, and the enhanced CuREV+ dataset, we aim to assess whether reformulated reviews lead to more efficient automation of the comment generation task.

\paragraph{\textbf{Model and data selection}}
To ensure a fair comparison, we selected a subset of $20,000$ comments from each dataset version (original, CuREV, CuREV+), such that each original review comment \( r_i \) is paired with its reformulated counterpart \( r'_i \) in CuREV and its further curated version in CuREV+. This selection strategy guarantees that any observed differences in model performance can be attributed to the quality of the data (\ie curation process) rather than differences in review content or model hyperparameters. We further split each subset into $75\%$ for training and $25\%$ for evaluation.

We selected \textit{DeepSeek-6.7B-Instruct}~\citep{deepseek-coder}, an LLM tailored for code-related tasks. 
Given a code change, the model was tasked with generating either the original, CuREV, or CuREV+ comment. To ensure consistency, we trained three separate models, one for each dataset version, using identical configurations.

\paragraph{\textbf{Experimental setup}}

This experiment aims to determine whether curated review comments improve the ability of LLMs to generate accurate review comments. For each dataset version (original, CuREV, CuREV+), we provided the model with code changes as input and tasked it with generating the corresponding review comment.

Each model was trained independently using the same configuration to ensure that observed performance differences could be attributed solely to the dataset quality, not to model or hyperparameter variations. The training was conducted on four \emph{NVIDIA RTX A5000 GPUs}, each with \emph{24GB} of memory. We used a batch size of $4$ and trained each model for $5$ epochs. To enable efficient, low-resource fine-tuning, we employed Low-Rank Adaptation (LoRA)~\citep{hu2021lora}, a parameter-efficient fine-tuning technique, configured with settings of $r = 16$, $\alpha = 32$, and $dropout = 0.05$. LoRA operates by decomposing the weight updates of a neural network into low-rank matrices, significantly reducing the number of parameters that require updating during fine-tuning~\citep{hu2021lora}, thus enhancing the overall efficiency of the training process. LoRA has been widely used in prior work to fine-tune LLMs for software engineering tasks~\citep{lu2023llama, weyssow2023exploring, hou2023large, silva2023repairllama}.

To evaluate the three produced models’ performance, we used the BLEU score~\citep{papineni2002bleu}, a standard metric, widely used in the literature, that measures the precision of n-grams in the generated text relative to the ground truth. BLEU is well-suited for assessing the correctness of generated comments, with higher scores indicating greater accuracy with real output.

\paragraph{\textbf{Results}}

The results are shown in \Table{tab:com_results}. The model trained on the original dataset obtained a BLEU score of $7.71$, compared to $11.26$ with CuREV. CuREV+ achieved $11.05$, slightly below CuREV but still markedly higher than the original dataset.

This outcome reflects a trade-off between performance and diversity. CuREV, composed entirely of reformulated comments, provides uniform phrasing that favors n-gram overlap and boosts BLEU scores. In contrast, CuREV+ blends reformulated with natural comments, introducing stylistic variation that slightly reduces n-gram alignment while still preserving strong learnability. The fact that CuREV+ remains close to CuREV in BLEU, while substantially outperforming the original dataset, suggests it strikes a valuable balance between accuracy in comment generation and stylistic diversity.

Moreover, the gains from CuREV may be partly inflated by repeated tokens—most notably the frequent use of ``consider'' at the start of comments—which artificially raises BLEU. CuREV+, by reducing such repetition and introducing more variety, may therefore offer a more realistic evaluation of a model’s ability to generalize beyond formulaic comment structures.

Overall, these findings confirm that curated review comments, whether CuREV or CuREV+, provide clearer, more direct guidance, enabling models to better capture intended messages. The consistently higher BLEU scores show that the curation process improves generalization, leading to more accurate review comments and facilitating more effective learning of review comments.

\begin{table}[!htbp]
\centering
\caption{Comparison of BLEU scores for DeepSeek-Coder-6.7B-Instruct trained on original, CuREV, and CuREV+ comments.}
\label{tab:com_results}
\begin{tabular}{@{}lccc@{}}
\toprule
& \textbf{Original Comments} & \textbf{CuREV} & \textbf{CuREV+} \\ 
\midrule
\textbf{BLEU} & $7.71$ & $11.26$ & $11.05$ \\ 
\bottomrule
\end{tabular}
\end{table}

\begin{center}
\begin{tcolorbox}[
    colback=orange!5!white,
    colframe=orange!75!black,
    left=3mm,
    overlay={\draw[orange!80!black, line width=3pt] (frame.south west)--(frame.north west);},
    title=\textbf{Answer to RQ4 (Impact on Comment Generation)},
    fonttitle=\bfseries,
    boxrule=0pt,
    arc=2pt,
    before skip=5pt,
    after skip=5pt
]
Curated review comments substantially improve performance in the comment generation task. 
BLEU increases from $7.71$ with the Original dataset to $11.26$ with CuREV, and remains strong with $11.05$ on CuREV+. 
While CuREV benefits from uniform reformulations that maximize n-gram overlap, CuREV+ introduces a richer mix of reformulated and natural comments. This stylistic diversity slightly reduces BLEU but offers a more realistic evaluation of the ability of the model to generalize beyond formulaic patterns.  Thus, CuREV+ achieves a valuable balance between performance and natural variation in review comments. 
\end{tcolorbox}
\end{center}


\subsection{A Comparative Analysis on the Usefulness of Curated Comments for Code Refinement}

In this section, we evaluate the usefulness of the curated comments compared to the original comments for code refinement. 
We conduct a comparative study to assess which version of the comment—original, CuREV, or CuREV+—guides the code refinement model to generate more accurate code changes.

\paragraph{\textbf{Model and data selection}}

We use \textit{DeepSeek-Coder-6.7B-Instruct}~\citep{deepseek-coder} as our code refinement model, applying it to the same selected subset of $20,000$ samples from each dataset version (original, CuREV, CuREV+), as in the previous experiment on comment generation (\Sect{subsec:model_data}). To ensure a fair comparison, each original review comment \( r_i \) is paired with its reformulated counterpart \( r'_i \) in CuREV and its enhanced version in CuREV+, preserving the same review context across all datasets. The model configurations remain identical for all datasets, ensuring that any observed differences are attributable solely to the quality of the review comments rather than model variations.

\paragraph{\textbf{Experimental setup}}

For each dataset version, we provided the code refinement model with the original code diff, the old file, and the review comment (either original, CuREV, or CuREV+) as context, and prompted it to generate a code diff that accurately implements the specified changes.

We used the LLM directly for inference, as its extensive training on diverse code-related tasks equips it with the capabilities needed to effectively automate the code refinement task. The experiment was run three times, once for each dataset version.

To evaluate the accuracy of the generated code diffs, we employed two metrics: 
\begin{itemize} 
    \item \textbf{CodeBLEU}: This metric measures the similarity of the generated code diff to the expected code diff, combining n-gram match, weighted n-gram match, AST match, and data-flow match scores~\citep{ren2020codebleu}. 
    \item \textbf{Exact Match (EM)}: This metric calculates the number of generated code diffs that exactly match the expected code diff. 
\end{itemize}

Each experiment was conducted using identical model configurations for both dataset versions to ensure that any observed performance differences could be attributed solely to the quality of the review comments rather than model parameter variations.

\paragraph{\textbf{Results}}
The results, presented in \Table{tab:ref_results}, show a clear progression across dataset versions. 
With the \textbf{original comments}, the model achieved a \emph{CodeBLEU} score of $0.36$ and an \emph{EM} of $408$. 
Training on \textbf{CuREV} yielded notable improvements, reaching $0.44$ CodeBLEU and $445$ EM. 
The best performance was observed with \textbf{CuREV+}, where the model achieved a \emph{CodeBLEU} score of $0.49$ and an \emph{EM} of $463$. 

These findings suggest that the curated comments offer more precise guidance, enabling the model to generate more accurate code changes that are closer to the ground truth.

While CuREV enhances clarity and structure through systematic reformulation, CuREV+ further strengthens guidance by blending reformulated with high-quality natural comments. 
This combination provides both precision, clarity, conciseness, and stylistic diversity, enabling the model to more effectively interpret review instructions and generate code changes that align with the ground truth.

\begin{table}[h!]
\centering
\caption{Comparison of DeepSeek-Coder-6.7B-Instruct's effectiveness for code refinement using original, CuREV, and CuREV+ review comments.}
\label{tab:ref_results}
\begin{tabular}{@{}lccc@{}}
\toprule
\textbf{Dataset Version} & \textbf{CodeBLEU} & \textbf{Exact Match} \\ 
\midrule
Original Comments & $0.36$ & $408$ \\
CuREV             & $0.44$ & $445$ \\
CuREV+            & $\textbf{0.49}$ & $\textbf{463}$ \\
\bottomrule
\end{tabular}
\end{table}

This experiment shows that both curated datasets significantly improve the utility of review comments for code refinement, with CuREV+ providing the strongest gains in terms of \emph{CodeBLEU} and \emph{EM}.. 
By reducing ambiguities and promoting richer, more natural guidance, CuREV+ enables the model to produce code changes that are not only more accurate but also better aligned with realistic review practices.

\begin{center}
\begin{tcolorbox}[
    colback=black!5!white,
    colframe=black!75!black,
    left=3mm,
    overlay={\draw[black!80!black, line width=3pt] (frame.south west)--(frame.north west);},
    title=\textbf{Answer to RQ5 (Usefulness for Code Refinement)},
    fonttitle=\bfseries,
    boxrule=0pt,
    arc=2pt,
    before skip=5pt,
    after skip=5pt
]
Curated comments strongly enhance code refinement performance. 
BLEU improves from $0.36$ with the Original dataset to $0.44$ with CuREV and further to $0.49$ with CuREV+. 
Exact Match follows the same trend ($408 \rightarrow 445 \rightarrow 463$). 
This demonstrates that CuREV+ offers the best trade-off, combining clarity from reformulations with stylistic diversity and conciseness from natural comments, thereby producing the most effective guidance for automated code refinement.
\end{tcolorbox}
\end{center}

\section{Threats to Validity}
\label{sec:threats}

The evaluation results have shown that our proposed methodology is effective in curating review comments and improving the performance of downstream tasks (\ie comment generation and code refinement). However, certain threats may limit the validity of these evaluation results.

A primary threat pertains to the nature of the data, specifically the review comments, which may contain noise and potentially non-English or misspelled words. We have mitigated this by employing \emph{Llama-3.1-70B}, an LLM that utilizes Byte-Pair Encoding~\citep{sennrich2015neural}, a subword-based tokenization algorithm. This algorithm breaks unseen words into several frequently seen sub-words that can be effectively processed by the model.

Another concern relates to the reliability of using an LLM as a judge, as LLM-generated judgments may not always match the accuracy of human evaluations. However, employing a bigger LLM, specifically \emph{Llama-3.1-70B}, helps mitigate this issue due to its advanced capabilities in producing accurate judgments. Prior research supports this hypothesis, showing that highly capable LLMs align closely with human assessments, often achieving agreement rates comparable to those between human evaluators~\citep{zheng2024judging, li2023alpacaeval}. To further validate the reliability of LLM judgments in our study, we conducted a thorough sanity check on $100$ manually assessed samples. The obtained Cohen's kappa scores indicated strong agreement rates between human and LLM evaluations, reinforcing the reliability of \emph{Llama-3.1-70B} judgments.

\subsection{Construct and External Validity}
One notable construct validity threat stems from our restriction to a single, albeit large-scale, dataset \citep{li2022automating}. Additionally, the vast majority of our extracted repositories comprise English-centric code base interactions, thereby our findings are limited to English-language code reviews. From an algorithmic evaluation perspective, while our zero-shot and fine-tuned analyses reliably emulate downstream performance, our methodology does not currently include an extensive human-in-the-loop developer perception study for the final curated comments, which we leave for future exploration. Finally, our 100-sample LLM validation check leveraged random sampling; a rigorously stratified sampling approach across independent programming languages could further strengthen the generalization limits of these evaluations.

\section{Related Work}
\label{sec:related}

To assist developers in code review, various techniques and tools have been proposed to automate review comment generation.
Early approaches relied on information retrieval methods. For instance, Hong \etal~\citep{hong2022commentfinder} introduced CommentFinder, an approach that retrieves relevant past review comments for new code changes. Similarly, Gupta et al.~\citep{gupta2018intelligent} developed DeepCodeReviewer (DCR), an LSTM-based model that predicts relevant reviews by ranking them based on code similarity. Although these methods effectively recommend existing comments, their limitation lies in the inability to generate new comments for unseen code.

More advanced solutions leverage language models. Tufano \etal~\citep{tufan2021towards, tufano2022using} used the T5 transformer, pre-trained with a masked language modeling task, and fine-tuned to generate review comments for Java code. Building on this, Li \etal~\citep{li2022automating} introduced a CodeT5 model pre-trained on tasks specifically designed for code review, such as quality estimation and comment generation. This demonstrated significant advancements in handling multilingual datasets and downstream tasks, including comment generation and code refinement.

Recent work aimed to enhance the performance of the finetuning-based approaches. Sghaier \etal~\citep{sghaier2024improving} proposed DISCOREV, an approach that incorporates cross-task knowledge distillation, connecting quality estimation, comment generation, and code refinement. This technique uses a cascade of models where each task informs the fine-tuning of the next, showing improvements over previous models.

Although these approaches have shown promising results, none of the existing works in the literature examined the quality of code review datasets or implemented preprocessing techniques to curate the data. Instead, most efforts focused on the fine-tuning phase of pre-trained language models. This leaves a significant gap in addressing the limitations of existing raw datasets. Our work proposes curating a code review dataset to improve the automation of code review tasks, overcoming the challenges posed by noisy and unrefined data.

Shi et al.~\citep{shi2022we} investigate the impact of data preprocessing on code summarization tasks by analyzing four widely-used benchmark datasets. They identify several types of noisy data, such as incomplete comments, auto-generated code, and over-splitting of identifiers, which affect model performance. To address this, they propose an automated cleaning tool based on heuristic rules, which detects and removes noisy data. Their experiments show that filtering out noisy data significantly improves the performance of various code summarization models.

While this approach demonstrated the importance of dataset quality for code summarization, the employed rule-based approach focuses on syntactic and formatting issues (e.g., HTML tags, url format) and does not address semantic or domain-specific challenges inherent to data. Our work applies a more sophisticated curation pipeline to code review data by employing LLM to also handle semantic issues and contextual inconsistencies, while establishing criteria for review comment quality (e.g., clarity, conciseness, civility) that are specific to code reviews.

Our approach shares high-level architectural overlaps with tools like AR-miner \citep{chen2014ar}, which classifies app reviews into informative categories via topic modeling. However, our evaluation framework deviates by injecting code-aware dimensions and multi-label quality tracking (Type, Nature, Civility) rather than relying on a strictly binary informative split. 
Similarly, while conventional emotion mining paradigms in software engineering effectively map developer communications to sentiment polarities, our civility assessment extends beyond simple token-level sentiment. For instance, emotion mining often fails to flag passive-aggressive sarcasm masked by neutral sentiment, whereas our multi-dimensional evaluation accurately categorizes unconstructive, unprofessional dialogue.

\section{Conclusion}
\label{sec:conclusion}

In this work, we proposed a systematic methodology for enhancing the quality of code review datasets. We first introduced an evaluation framework for assessing review comments and identified widespread issues in existing datasets, including irrelevance, incivility, lack of clarity and conciseness, and stylistic homogeneity. To mitigate these issues, we designed two different curation pipelines. The first, \emph{CuREV}, reformulates all relevant comments to maximize clarity, conciseness, and civility. The second, \emph{CuREV+}, selectively reformulates only low-quality comments while preserving high-quality human-written ones, using in-context exemplars in reformulation to inject stylistic diversity.

Our comparative evaluation across three datasets (original, CuREV, and CuREV+) demonstrates the effectiveness of our proposed curation pipelines. On the comment generation task, both CuREV and CuREV+ outperform the original dataset, with CuREV reaching the highest BLEU due to its consistent phrasing, and CuREV+ achieving a slightly lower score but offering greater stylistic richness. On the code refinement task, CuREV+ yields the strongest results overall, with improvements in both \emph{CodeBLEU} and \emph{Exact Match}, confirming that the blend of reformulated and authentic comments provides models with clearer and more contextually useful guidance. Moreover, our diversity analysis shows that CuREV+, unlike CuREV, avoids excessive repetition of directive tokens and instead reflects a more human-like variety of styles and expressions.

These findings suggest that while systematic reformulation (CuREV) can enhance clarity and improve short-term model learning, selective and high-quality exemplars-guided reformulation (CuREV+) provides a stronger long-term balance between performance, naturalness, and diversity.

In future work, this work opens new research opportunities in curating datasets for software engineering tasks. Future directions include leveraging ensembles of LLMs as ``juries'' for more robust multi-perspective curation, exploring active learning strategies for prioritizing comments, and studying the impact of diversity on developer trust and adoption of automated review systems.


Looking forward, there are several imperative avenues for future research. While our extensive automated analyses (BLEU, BERTScore) firmly establish semantic equivalence and refinement quality, conducting an exhaustive human-in-the-loop developer perception study is critical. Furthermore, evaluating alternative LLMs during the actual \emph{reformulation} phase (beyond merely using multi-models exclusively as judges) remains an important frontier for maximizing stylistic authenticity.


\section*{Data Availability}
The curated datasets (CuREV and CuREV+) and the replication packages are available on HuggingFace~\citep{curev_hf, curev+_hf}, GitHub~\citep{curev_gh, curev+_gh}, and Zenodo~\citep{curev_zenodo, curev+_zenodo}.

\section*{Declarations}

\subsection*{Funding}
Not applicable. The authors did not receive any funding for this research.

\subsection*{Ethical Approval}
This study does not involve human participants or animals.

\subsection*{Informed Consent}
Not applicable. No human subjects were involved in this study.

\subsection*{Author Contributions}
-- Oussama Ben Sghaier: Conceptualization, Dataset Curation, Methodology, Experiments, Writing – Original Draft. \\
-- Martin Weyssow: Methodology, Data Validation, Writing – Review \& Editing. \\
-- Houari Sahraoui: Supervision, Conceptual Guidance, Research Direction, Writing – Review \& Editing.

\subsection*{Data Availability Statement}
The curated datasets (CuREV and CuREV+) and the replication packages are available on HuggingFace~\citep{curev_hf, curev+_hf}, GitHub~\citep{curev_gh, curev+_gh}, and Zenodo~\citep{curev_zenodo, curev+_zenodo}.

\subsection*{Conflict of Interest}
The authors declare that they have no known competing interests or personal relationships that could have appeared to influence the work reported in this article.

\subsection*{Clinical Trial Number}
Not applicable.


\bibliographystyle{spbasic}      
\bibliography{references}   

@article{sghaier2024improving,
  title={Improving the Learning of Code Review Successive Tasks with Cross-Task Knowledge Distillation},
  author={Sghaier, Oussama Ben and Sahraoui, Houari},
  journal={arXiv preprint arXiv:2402.02063},
  year={2024}
}

@inproceedings{lu2023llama,
  title={LLaMA-Reviewer: Advancing code review automation with large language models through parameter-efficient fine-tuning},
  author={Lu, Junyi and Yu, Lei and Li, Xiaojia and Yang, Li and Zuo, Chun},
  booktitle={2023 IEEE 34th International Symposium on Software Reliability Engineering (ISSRE)},
  pages={647--658},
  year={2023},
  organization={IEEE}
}

@article{silva2023repairllama,
  title={Repairllama: Efficient representations and fine-tuned adapters for program repair},
  author={Silva, Andr{\'e} and Fang, Sen and Monperrus, Martin},
  journal={arXiv preprint arXiv:2312.15698},
  year={2023}
}

@article{hou2023large,
  title={Large language models for software engineering: A systematic literature review},
  author={Hou, Xinyi and Zhao, Yanjie and Liu, Yue and Yang, Zhou and Wang, Kailong and Li, Li and Luo, Xiapu and Lo, David and Grundy, John and Wang, Haoyu},
  journal={ACM Transactions on Software Engineering and Methodology},
  year={2023},
  publisher={ACM New York, NY}
}

@article{weyssow2023exploring,
  title={Exploring parameter-efficient fine-tuning techniques for code generation with large language models},
  author={Weyssow, Martin and Zhou, Xin and Kim, Kisub and Lo, David and Sahraoui, Houari},
  journal={arXiv preprint arXiv:2308.10462},
  year={2023}
}

@article{chang2024survey,
  title={A survey on evaluation of large language models},
  author={Chang, Yupeng and Wang, Xu and Wang, Jindong and Wu, Yuan and Yang, Linyi and Zhu, Kaijie and Chen, Hao and Yi, Xiaoyuan and Wang, Cunxiang and Wang, Yidong and others},
  journal={ACM Transactions on Intelligent Systems and Technology},
  volume={15},
  number={3},
  pages={1--45},
  year={2024},
  publisher={ACM New York, NY}
}

@article{zhuo2023ice,
  title={ICE-Score: Instructing Large Language Models to Evaluate Code},
  author={Zhuo, Terry Yue},
  journal={arXiv preprint arXiv:2304.14317},
  year={2023}
}

@article{zheng2023judging,
  title={Judging llm-as-a-judge with mt-bench and chatbot arena},
  author={Zheng, Lianmin and Chiang, Wei-Lin and Sheng, Ying and Zhuang, Siyuan and Wu, Zhanghao and Zhuang, Yonghao and Lin, Zi and Li, Zhuohan and Li, Dacheng and Xing, Eric and others},
  journal={Advances in Neural Information Processing Systems},
  volume={36},
  pages={46595--46623},
  year={2023}
}

@INPROCEEDINGS{bacchelli2013expectations,  author={Bacchelli, Alberto and Bird, Christian},  booktitle={2013 35th International Conference on Software Engineering (ICSE)},   title={Expectations, outcomes, and challenges of modern code review},   year={2013},  volume={},  number={},  pages={712-721},  doi={10.1109/ICSE.2013.6606617}}

@inproceedings{mcintosh2014impact,
  title={The impact of code review coverage and code review participation on software quality: A case study of the {Qt}, {VTK}, and {ITK} projects},
  author={McIntosh, Shane and Kamei, Yasutaka and Adams, Bram and Hassan, Ahmed E},
  booktitle={11th working conference on mining software repositories},
  pages={192--201},
  year={2014}
}

@article{mcintosh2016empirical,
  title={An empirical study of the impact of modern code review practices on software quality},
  author={McIntosh, Shane and Kamei, Yasutaka and Adams, Bram and Hassan, Ahmed E},
  journal={Empirical Software Engineering},
  volume={21},
  number={5},
  pages={2146--2189},
  year={2016},
  publisher={Springer}
}

@article{ackerman1989software,
  title={Software inspections: an effective verification process},
  author={Ackerman, A. Frank and Buchwald, Lynne S. and Lewski, Frank H.},
  journal={IEEE software},
  volume={6},
  number={3},
  pages={31--36},
  year={1989},
  publisher={IEEE}
}

@inproceedings{gupta2018intelligent,
  title={Intelligent code reviews using deep learning},
  author={Gupta, Anshul and Sundaresan, Neel},
  booktitle={Proceedings of the 24th ACM SIGKDD International Conference on Knowledge Discovery and Data Mining (KDD’18) Deep Learning Day},
  year={2018}
}

@article{tufano2022using,
  title={Using pre-trained models to boost code review automation},
  author={Tufano, Rosalia and Masiero, Simone and Mastropaolo, Antonio and Pascarella, Luca and Poshyvanyk, Denys and Bavota, Gabriele},
  journal={arXiv preprint arXiv:2201.06850},
  year={2022}
}

@inproceedings{tufan2021towards,
  title={Towards automating code review activities},
  author={Tufano, Rosalia and Pascarella, Luca and Tufanoy, Michele and Poshyvanykz, Denys and Bavota, Gabriele},
  booktitle={2021 IEEE/ACM 43rd International Conference on Software Engineering (ICSE)},
  pages={163--174},
  year={2021},
}

@inproceedings{bavota2015four,
  title={Four eyes are better than two: On the impact of code reviews on software quality},
  author={Bavota, Gabriele and Russo, Barbara},
  booktitle={2015 IEEE International Conference on Software Maintenance and Evolution (ICSME)},
  pages={81--90},
  year={2015},
  organization={IEEE}
}

@incollection{fagan2002design,
  title={Design and code inspections to reduce errors in program development},
  author={Fagan, Michael},
  booktitle={Software pioneers},
  pages={575--607},
  year={2002},
  publisher={Springer}
}

@inproceedings{morales2015code,
  title={Do code review practices impact design quality? A case study of the {Qt}, {VTK}, and {ITK} projects},
  author={Morales, Rodrigo and McIntosh, Shane and Khomh, Foutse},
  booktitle={2015 IEEE 22nd international conference on software analysis, evolution, and reengineering (SANER)},
  pages={171--180},
  year={2015},
  organization={IEEE}
}

@article{sennrich2015neural,
  title={Neural machine translation of rare words with subword units},
  author={Sennrich, Rico and Haddow, Barry and Birch, Alexandra},
  journal={arXiv preprint arXiv:1508.07909},
  year={2015}
}

@inproceedings{li2022automating,
  title={Automating code review activities by large-scale pre-training},
  author={Li, Zhiyu and Lu, Shuai and Guo, Daya et al.},
  booktitle={Proceedings of the 30th ACM Joint European Software Engineering Conference and Symposium on the Foundations of Software Engineering},
  pages={1035--1047},
  year={2022}
}

@inproceedings{papineni2002bleu,
  title={Bleu: a method for automatic evaluation of machine translation},
  author={Papineni, Kishore and Roukos, Salim and Ward, Todd and Zhu, Wei-Jing},
  booktitle={Proceedings of the 40th annual meeting of the Association for Computational Linguistics},
  pages={311--318},
  year={2002}
}

@inproceedings{sghaier2023multi,
  title={A Multi-Step Learning Approach to Assist Code Review},
  author={Sghaier, Oussama Ben and Sahraoui, Houari},
  booktitle={2023 IEEE 23rd International Conference on Software Analysis, Evolution, and Reengineering (SANER)},
  year={2023},
  organization={IEEE}
}

@article{ren2020codebleu,
  title={Codebleu: a method for automatic evaluation of code synthesis},
  author={Ren, Shuo and Guo, Daya and Lu, Shuai and Zhou, Long and Liu, Shujie and Tang, Duyu and Sundaresan, Neel and Zhou, Ming and Blanco, Ambrosio and Ma, Shuai},
  journal={arXiv preprint arXiv:2009.10297},
  year={2020}
}

@inproceedings{hong2022commentfinder,
  title={Commentfinder: a simpler, faster, more accurate code review comments recommendation},
  author={Hong, Yang and Tantithamthavorn, Chakkrit and Thongtanunam, Patanamon and Aleti, Aldeida},
  booktitle={Proceedings of the 30th ACM joint European software engineering conference and symposium on the foundations of software engineering},
  pages={507--519},
  year={2022}
}

@misc{li2023alpacaeval,
  title={Alpacaeval: An automatic evaluator of instruction-following models},
  author={Li, Xuechen and Zhang, Tianyi and Dubois, Yann and Taori, Rohan and Gulrajani, Ishaan and Guestrin, Carlos and Liang, Percy and Hashimoto, Tatsunori B},
  year={2023}
}

@article{zheng2024judging,
  title={Judging llm-as-a-judge with mt-bench and chatbot arena},
  author={Zheng, Lianmin and Chiang, Wei-Lin and Sheng, Ying and Zhuang, Siyuan and Wu, Zhanghao and Zhuang, Yonghao and Lin, Zi and Li, Zhuohan and Li, Dacheng and Xing, Eric and others},
  journal={Advances in Neural Information Processing Systems},
  volume={36},
  year={2024}
}

@article{hu2021lora,
  title={Lora: Low-rank adaptation of large language models},
  author={Hu, Edward J and Shen, Yelong and Wallis, Phillip and Allen-Zhu, Zeyuan and Li, Yuanzhi and Wang, Shean and Wang, Lu and Chen, Weizhu},
  journal={arXiv preprint arXiv:2106.09685},
  year={2021}
}

@article{weyssow2024codeultrafeedback,
  title={CodeUltraFeedback: An LLM-as-a-Judge Dataset for Aligning Large Language Models to Coding Preferences},
  author={Weyssow, Martin and Kamanda, Aton and Sahraoui, Houari},
  journal={arXiv preprint arXiv:2403.09032},
  year={2024}
}

@inproceedings{tufano2021towards,
  title={Towards automating code review activities},
  author={Tufano, Rosalia and Pascarella, Luca and Tufano, Michele and Poshyvanyk, Denys and Bavota, Gabriele},
  booktitle={2021 IEEE/ACM 43rd International Conference on Software Engineering (ICSE)},
  pages={163--174},
  year={2021},
  organization={IEEE}
}

@inproceedings{li2022auger,
  title={AUGER: automatically generating review comments with pre-training models},
  author={Li, Lingwei and Yang, Li and Jiang, Huaxi and Yan, Jun and Luo, Tiejian and Hua, Zihan and Liang, Geng and Zuo, Chun},
  booktitle={Proceedings of the 30th ACM Joint European Software Engineering Conference and Symposium on the Foundations of Software Engineering},
  pages={1009--1021},
  year={2022}
}

@article{ben2024improving,
  title={Improving the learning of code review successive tasks with cross-task knowledge distillation},
  author={Ben Sghaier, Oussama and Sahraoui, Houari},
  journal={Proceedings of the ACM on Software Engineering},
  volume={1},
  number={FSE},
  pages={1086--1106},
  year={2024},
  publisher={ACM New York, NY, USA}
}

@misc{deepseek-coder,
  author = {Guo, Daya and Zhu, Qihao and Yang, Dejian et al.},
  title = {DeepSeek-Coder: When the Large Language Model Meets Programming -- The Rise of Code Intelligence},
  journal = {CoRR},
  volume = {abs/2401.14196},
  year = {2024},
  url = {https://arxiv.org/abs/2401.14196},
}

@article{mchugh2012interrater,
  title={Interrater reliability: the kappa statistic},
  author={McHugh, Mary L},
  journal={Biochemia medica},
  volume={22},
  number={3},
  pages={276--282},
  year={2012},
  publisher={Medicinska naklada}
}

@article{tufano2024code,
  title={Code review automation: strengths and weaknesses of the state of the art},
  author={Tufano, Rosalia and Dabi{\'c}, Ozren and Mastropaolo, Antonio and Ciniselli, Matteo and Bavota, Gabriele},
  journal={IEEE Transactions on Software Engineering},
  year={2024},
  publisher={IEEE}
}

@article{rahman2024words,
  title={Do Words Have Power? Understanding and Fostering Civility in Code Review Discussion},
  author={Rahman, Md Shamimur and Codabux, Zadia and Roy, Chanchal K},
  journal={Proceedings of the ACM on Software Engineering},
  volume={1},
  number={FSE},
  pages={1632--1655},
  year={2024},
  publisher={ACM New York, NY, USA}
}

@article{rani2023decade,
  title={A decade of code comment quality assessment: A systematic literature review},
  author={Rani, Pooja and Blasi, Arianna and Stulova, Nataliia and Panichella, Sebastiano and Gorla, Alessandra and Nierstrasz, Oscar},
  journal={Journal of Systems and Software},
  volume={195},
  pages={111515},
  year={2023},
  publisher={Elsevier}
}

@inproceedings{haouari2011good,
  title={How good is your comment? a study of comments in java programs},
  author={Haouari, Dorsaf and Sahraoui, Houari and Langlais, Philippe},
  booktitle={2011 International symposium on empirical software engineering and measurement},
  pages={137--146},
  year={2011},
  organization={IEEE}
}

@article{sghaier2023unity,
  title={Unity is Strength: Cross-Task Knowledge Distillation to Improve Code Review Generation},
  author={Sghaier, Oussama Ben and Maes, Lucas and Sahraoui, Houari},
  journal={arXiv preprint arXiv:2309.03362},
  year={2023}
}

@inproceedings{terryn2024exploratory,
  title={Exploratory Study on the Impact of English Bias of Generative Large Language Models in Dutch and French},
  author={Terryn, Ayla Rigouts and de Lhoneux, Miryam},
  booktitle={Proceedings of the Fourth Workshop on Human Evaluation of NLP Systems (HumEval)@ LREC-COLING 2024},
  pages={12--27},
  year={2024}
}

@inproceedings{shi2022we,
  title={Are we building on the rock? on the importance of data preprocessing for code summarization},
  author={Shi, Lin and Mu, Fangwen and Chen, Xiao and Wang, Song and Wang, Junjie and Yang, Ye and Li, Ge and Xia, Xin and Wang, Qing},
  booktitle={Proceedings of the 30th ACM Joint European Software Engineering Conference and Symposium on the Foundations of Software Engineering},
  pages={107--119},
  year={2022}
}

@article{zhou2021evaluating,
  title={Evaluating the quality of machine learning explanations: A survey on methods and metrics},
  author={Zhou, Jianlong and Gandomi, Amir H and Chen, Fang and Holzinger, Andreas},
  journal={Electronics},
  volume={10},
  number={5},
  pages={593},
  year={2021},
  publisher={MDPI}
}

@article{li2015diversity,
  title={A diversity-promoting objective function for neural conversation models},
  author={Li, Jiwei and Galley, Michel and Brockett, Chris and Gao, Jianfeng and Dolan, Bill},
  journal={arXiv preprint arXiv:1510.03055},
  year={2015}
}

@inproceedings{zhu2018texygen,
  title={Texygen: A Benchmarking Platform for Text Generation Models},
  author={Zhu, Yaoming and Lu, Sidi and Zheng, Lei and Guo, Jiaxian and Zhang, Weinan and Wang, Jun and Yu, Yong},
  booktitle={The 41st International ACM SIGIR Conference on Research and Development in Information Retrieval (SIGIR '18)},
  pages={1097--1100},
  year={2018},
  publisher={ACM}
}

@inproceedings{sghaier2025harnessing,
  title={Harnessing Large Language Models for Curated Code Reviews},
  author={Sghaier, Oussama Ben and Weyssow, Martin and Sahraoui, Houari},
  booktitle={2025 IEEE/ACM 22nd International Conference on Mining Software Repositories (MSR)},
  pages={187--198},
  year={2025},
  organization={IEEE}
}

@inproceedings{lu2025deepcrceval,
  title={Deepcrceval: Revisiting the evaluation of code review comment generation},
  author={Lu, Junyi and Li, Xiaojia and Hua, Zihan and Yu, Lei and Cheng, Shiqi and Yang, Li and Zhang, Fengjun and Zuo, Chun},
  booktitle={International Conference on Fundamental Approaches to Software Engineering},
  pages={43--64},
  year={2025},
  organization={Springer Nature Switzerland Cham}
}

@inproceedings{sadowski2018modern,
  title={Modern code review: a case study at google},
  author={Sadowski, Caitlin and S{\"o}derberg, Emma and Church, Luke and Sipko, Michal and Bacchelli, Alberto},
  booktitle={Proceedings of the 40th international conference on software engineering: Software engineering in practice},
  pages={181--190},
  year={2018}
}

@misc{curev_zenodo,
  title = {CuREV Replication package},
  key = {Data and Models},
  howpublished = {\url{https://zenodo.org/records/14812107}},
  year={2025}
}

@misc{curev_hf,
  title = {CuREV HF Dataset},
  key = {CuREV - HF Dataset},
  howpublished = {\url{https://huggingface.co/datasets/OussamaBS/CuREV}},
  year={2025}
}

@misc{curev_gh,
  title = {CuREV Repository},
  key = {CuREV Repository},
  howpublished = {\url{https://github.com/OussamaSghaier/CuREV}},
  year={2025}
}

@misc{curev+_zenodo,
  title = {CuREV+ - Replication package},
  key = {CuREV+ - Replication package},
  howpublished = {\url{https://zenodo.org/records/17337508}},
  year={2025}
}

@misc{curev+_hf,
  title = {CuREV+ HF Dataset},
  key = {CuREV+ HF Dataset},
  howpublished = {\url{https://huggingface.co/datasets/OussamaBS/CuREV-plus}},
  year={2025}
}

@misc{curev+_gh,
  title = {CuREV+ Repository},
  key = {CuREV+ Repository},
  howpublished = {\url{https://github.com/OussamaSghaier/CuREV-plus}},
  year={2025}
}

@inproceedings{zhang2019bertscore,
  title={BERTScore: Evaluating Text Generation with BERT},
  author={Zhang, Tianyi and Kishore, Varsha and Wu, Felix and Weinberger, Kilian Q and Artzi, Yoav},
  booktitle={International Conference on Learning Representations},
  year={2020}
}

@inproceedings{chen2014ar,
  title={AR-miner: mining informative reviews for developers from mobile app marketplace},
  author={Chen, Ning and Lin, Jialiu and Hoi, Steven CH and Xiao, Xiaokui and Zhang, Boshen},
  booktitle={Proceedings of the 36th international conference on software engineering},
  pages={767--778},
  year={2014}
}

@article{hindle2012naturalness,
  title={On the Naturalness of Software (ICSE'12)},
  author={Hindle, Abram and Barr, Earl T and Su, Zhendong and Gabel, Mark and Devanbu, Premkumar},
  journal={ICSE’12},
  year={2012}
}

\end{document}